\documentclass[aps,prd,twocolumn,showpacs,amsmath,amssymb]{revtex4-1}
\usepackage{epsfig}
\usepackage{graphicx}
\usepackage{dcolumn}
\usepackage{bm}
\usepackage{overpic}
\usepackage{subfigure}
\usepackage{float}
\usepackage{color}
\usepackage{amsmath}
\usepackage{mathcomp}
\usepackage{mathrsfs}
\usepackage{multirow}
\usepackage{rotating}

\begin{document}
\normalsize
\parskip=5pt plus 1pt minus 1pt

\title{ \boldmath Amplitude analysis of $D^{+} \rightarrow K_{S}^{0} \pi^{+} \pi^{+} \pi^{-}$ }
\vspace{-1cm}

\author{
   \begin{small}
    \begin{center}
      M.~Ablikim$^{1}$, M.~N.~Achasov$^{10,d}$, S. ~Ahmed$^{15}$, M.~Albrecht$^{4}$, M.~Alekseev$^{55A,55C}$, A.~Amoroso$^{55A,55C}$, F.~F.~An$^{1}$, Q.~An$^{42,52}$, Y.~Bai$^{41}$, O.~Bakina$^{27}$
% ...
         \\
         \vspace{0.2cm}
   (BESIII Collaboration)\\
\vspace{0.2cm} {\it
$^{1}$ Institute of High Energy Physics, Beijing 100049, People's Republic of China\\
% ...
$^{4}$ Bochum Ruhr-University, D-44780 Bochum, Germany\\
% ...
$^{10}$ G.I. Budker Institute of Nuclear Physics SB RAS (BINP), Novosibirsk 630090, Russia\\
% ...
$^{15}$ Helmholtz Institute Mainz, Johann-Joachim-Becher-Weg 45, D-55099 Mainz, Germany\\
% ...
$^{27}$ Joint Institute for Nuclear Research, 141980 Dubna, Moscow region, Russia\\
% ...
$^{41}$ Southeast University, Nanjing 211100, People's Republic of China\\
$^{42}$ State Key Laboratory of Particle Detection and Electronics, Beijing 100049, Hefei 230026, People's Republic of China\\
% ...
$^{52}$ University of Science and Technology of China, Hefei 230026, People's Republic of China\\
% ...
$^{55}$ (A)University of Turin, I-10125, Turin, Italy; (B)University of Eastern Piedmont, I-15121, Alessandria, Italy; (C)INFN, I-10125, Turin, Italy\\
% ...
$^{d}$ Also at the Novosibirsk State University, Novosibirsk, 630090, Russia\\
% ...
}\end{center}
\vspace{0.4cm}
\end{small}
}

\affiliation{}
\vspace{-4cm}
\date{\today}
\begin{abstract}
The decay $D^{+} \rightarrow K_{S}^{0} \pi^{+} \pi^{+} \pi^{-}$ is studied with an amplitude analysis
using a data set of 2.93${\mbox{\,fb}^{-1}}$ of $e^+e^+$ collisions at the  $\psi(3770)$ peak accumulated by the BESIII detector.
Intermediate states  and non-resonant components,
and their relative fractions and phases have been determined.  The significant amplitudes, which contribute to the model that best fits the data, are composed of 
 five quasi-two-body decays $ K_{S}^{0} a_{1}(1260)^{+}$,
$ \bar{K}_{1}(1270)^{0} \pi^{+}$
$ \bar{K}_{1}(1400)^{0} \pi^{+}$,
$ \bar{K}_{1}(1650)^{0} \pi^{+}$,
and $ \bar{K}(1460)^{0} \pi^{+}$,
a three-body decays $K_{S}^{0}\pi^{+}\rho^{0}$, as well as
a non-resonant component $ K_{S}^{0}\pi^{+}\pi^{+}\pi^{-}$.
The dominant amplitude  is $ K_{S}^{0} a_{1}(1260)^{+}$,
with a fit fraction of $(40.3\pm2.1\pm2.9)\%$, where the first and second uncertainties are statistical and systematic, respectively.
\end{abstract}
\maketitle

\section{Introduction}
\label{sec:introduction}

Hadronic decays of mesons with charm are an important tool for understanding the dynamics of the strong interaction in the low energy regime.
The amplitudes describing $D$ meson weak decays into
four-body final states are dominated by (quasi)-two-body processes, such as 
$D \rightarrow VP$, $D \rightarrow SP$, $D \rightarrow VV$, and $D \rightarrow AP$, where 
$P$, $V$, $S$, and $A$ denote pseudoscalar, vector, scalar, and axial-vector mesons, respectively.  
Final-state interactions can cause significant changes in decay rates and shifts in the phases of decay amplitudes. 
Experimental measurements can help to refine theoretical models of these phenomena~\cite{Cheng:2003bn,PRD81074031,HYCheng}.
Many measurements on $D \rightarrow PP$ and $D \rightarrow VP$ decays have been performed~\cite{PDG}.
However, there are only a few studies focusing on $D \rightarrow AP$ decays~\cite{PDG}.  
We have therefore measured $D \rightarrow AP$ decays via an amplitude analysis of 
the decay $D^{+} \rightarrow K_{S}^{0}\pi^{+}\pi^{+}\pi^{-}$ (the inclusion of charge conjugate reaction 
is implied throughout the paper),  which is expected to be dominated by $D^+ \rightarrow K_{S}^{0} a_{1}(1260)^{+}$. 
In addition, the measurements of the intermediate processes containing ${K}_{1}(1270)$ and ${K}_{1}(1400)$ will be 
helpful for understanding the mixture between these two axial-vector  kaons~\cite{HYCheng}.

In this paper, we present an amplitude analysis of the decay 
$D^{+} \rightarrow K_{S}^{0} \pi^{+} \pi^{+} \pi^{-}$ to study
the resonant substructures and non-resonant components, 
where the amplitude model is constructed using the covariant tensor formalism~\cite{Zou}.

\section{Detector and Data Sets}
\label{sec:Data Set and Event Seletion}
The data used in this analysis were accumulated with the  BESIII detector~\cite{detector}.  
The event sample is based on 2.93 ${\mbox{\,fb}^{-1}}$ of $e^+e^-$ collisions at the $\psi(3770)$ mass~\cite{datasample, datasample2}. 
At this energy, $D$ meson pairs are produced without any additional hadrons.
To suppress backgrounds from other charmed meson decays and continuum
 (QED and $q\bar{q}$) processes,  only the decay mode  $D^- \to K^+ \pi^- \pi^-$  is used to tag the $D^+D^-$ pairs. 
This provides a clean environment for selecting the decay $D^{+} \rightarrow K_{S}^{0} \pi^{+} \pi^{+} \pi^{-}$ (the signal side)
by requiring the $D^{-}\rightarrow K^{+} \pi^{-} \pi^{-}$ decay to be observed (the tag side).

The BESIII detector located at Beijing Electron 
Positron Collider~\cite{Yu:IPAC2016-TUYA01} is described in Ref.~\cite{detector}.
The geometrical acceptance of the BESIII detector is  93\% of the full solid angle.
Starting from the interaction point (IP), it consists of a main drift chamber (MDC),
a time-of-flight (TOF) system, a CsI(Tl) electromagnetic calorimeter, which are all enclosed in a superconducting solenoidal magnet
providing a 1.0~T magnetic field. The solenoid is supported by an
octagonal flux-return yoke with resistive plate counter muon
identifier modules interleaved with steel. 
The momentum resolution for charged tracks in the MDC is 0.5\% at a transverse momentum of 1 GeV$/c$.
The energy resolution for photon in EMC measurement is 2.5\% (5\%) in the barrel (end caps) region at 1 GeV.
The time resolution of the TOF barrel part is 68~ps, while that of the end cap part is 110~ps.

Monte Carlo (MC) simulations of the BESIII detector are based on
{\sc geant4}~\cite{sim}.
The production of $\psi(3770)$ is simulated with the {\sc kkmc}~\cite{KKMC} package,
taking into account the beam energy spread and the initial-state radiation (ISR).
The {\sc photos}~\cite{FSR} package is used to simulate the final-state radiation of charged particles.
The {\sc evtgen}~\cite{EvtGen} package is used to simulate the known decay modes with branching fractions (BFs) taken from
the Particle Data Group (PDG)~\cite{PDG}, and the remaining unknown decays are generated with the {\sc LundCharm} model~\cite{LundCharm}.
The MC sample referred to as ``generic MC'', including the processes of $\psi(3770)$ decays to $D\bar{D}$, non-$D\bar{D}$,
ISR production of low mass charmonium states and continuum processes, is used to
study the background contribution.
The effective luminosities of the generic MC samples correspond to at least 5 times the data sample luminosity. 
Two kind of MC samples with the decay chain of $\psi(3770) \rightarrow D^+ D^-$ with 
$D^{+} \rightarrow K_{S}^{0}\pi^{+}\pi^{+}\pi^{-}$ and $D^{-} \rightarrow K^{+}\pi^{-}\pi^{-}$ using different decay models are generated for the amplitude analysis, .
One sample, ``PHSP MC'', is generated with an uniform distribution in phase space
for the $D^{+} \rightarrow K_{S}^{0}\pi^{+}\pi^{+}\pi^{-}$ decay, which is used  to calculate the MC integrations. 
The other sample,``signal MC'' , is generated according to the results obtained in this analysis for the $D^{+} \rightarrow K_{S}^{0}\pi^{+}\pi^{+}\pi^{-}$ decay.
It is used to validate the fit performance, calculate the goodness of fit and estimate the detector efficiency. 

\section{Event Selection}
\label{selection}

Good charged tracks other than $K_{S}^{0}$ daughters are required to have a point of closest approach
to the IP within $10$ cm along the beam axis and
within $1$ cm in the plane perpendicular to the beam.
The polar angle $\theta$ between the track and the $e^+$ beam direction
is required to satisfy $|\cos \theta|<0.93$. 
Separation of charged kaons from charged pions is implemented by combining the
energy loss ($dE/dx$) in the MDC and the time-of-fight information from the TOF.
We calculate the probabilities $P(K)$ and $P(\pi)$ with the hypothesis of $K$ or
$\pi$, and require that $K$ candidates have $P(K) > P(\pi)$,
while $\pi$ candidates have $P(\pi) > P(K)$.
Tracks without particle identification (PID) information are rejected.
Furthermore, a vertex fit with the hypothesis that all tracks originate from the IP is performed, and the $\chi^2$ of the fit is required to be less than 100.

The $K_{S}^{0}$ candidates are reconstructed from a pair of oppositely charged
tracks which satisfy $|\cos \theta|<0.93$ and whose distances to the IP along 
the beam direction are within $20$ cm.
The two charged tracks are assumed to be a $\pi^+\pi^-$ pair without PID.  
In order to improve the signal-to-background ratio, the
decay vertex of the $\pi^+\pi^-$ pair is required to be more than two
standard deviations
away from the IP~\cite{Xu:2009zzg}, and their invariant mass is
required to be in the region  $[467.6,\,527.6]$ MeV$/c^{2}$.

\begin{figure*}[hbtp]
\begin{center}
\centering
\begin{minipage}[b]{0.35\textwidth}
\epsfig{width=0.98\textwidth,file=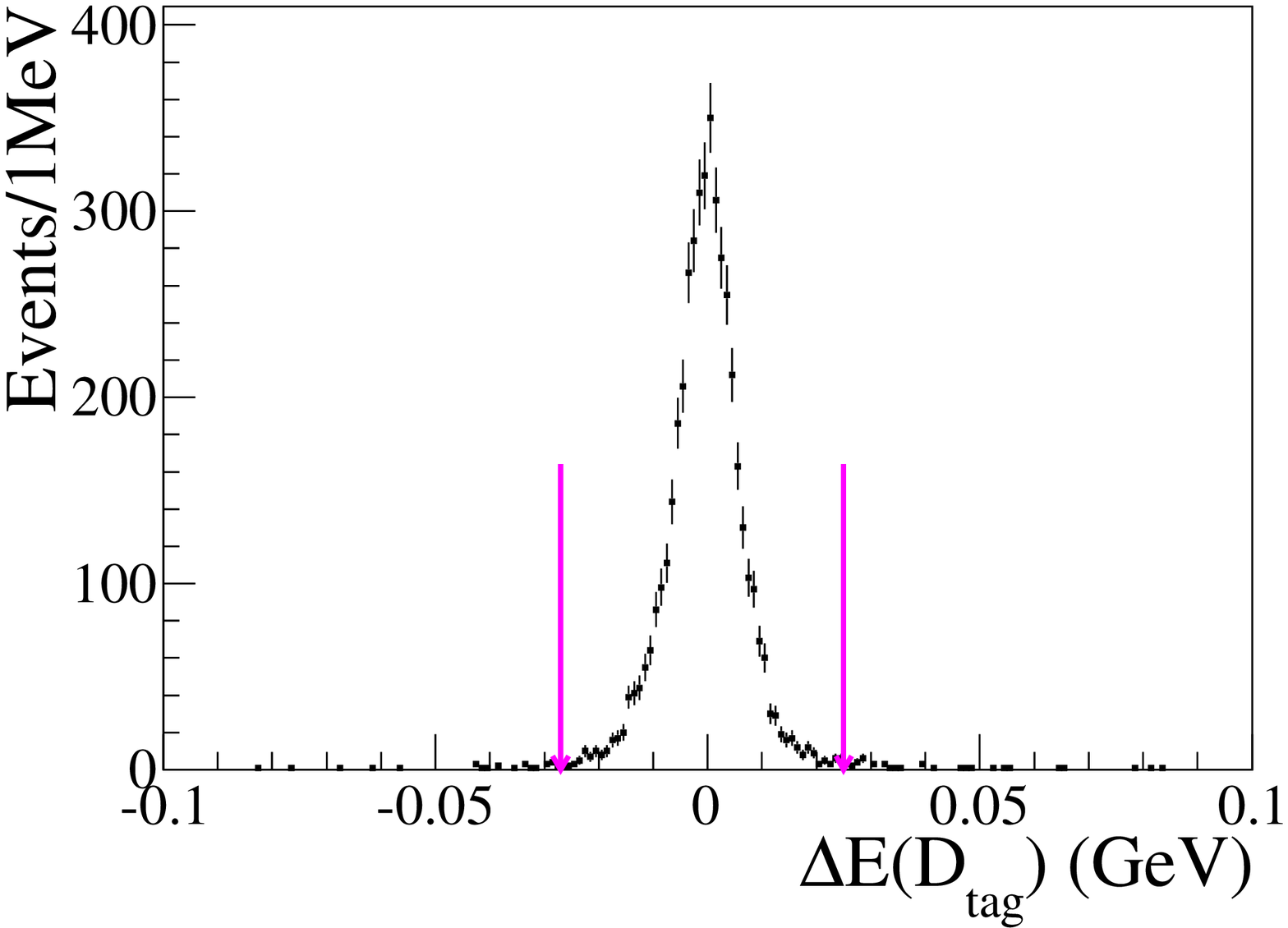}
\put(-130,100){(a)}
\end{minipage}
\begin{minipage}[b]{0.35\textwidth}
\epsfig{width=0.98\textwidth,file=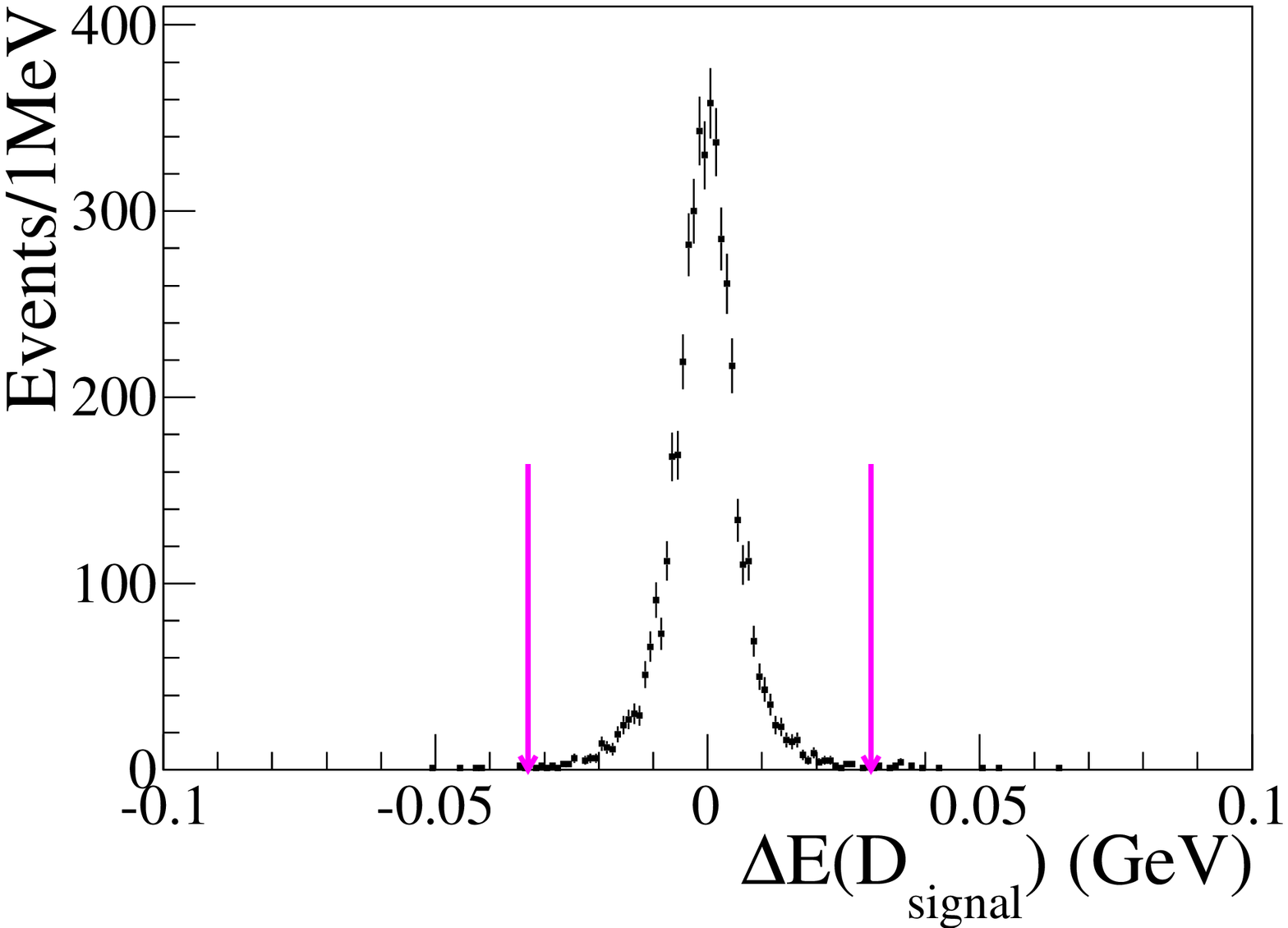}
\put(-130,100){(b)}
\end{minipage}
\begin{minipage}[b]{0.35\textwidth}
\epsfig{width=0.98\textwidth,file=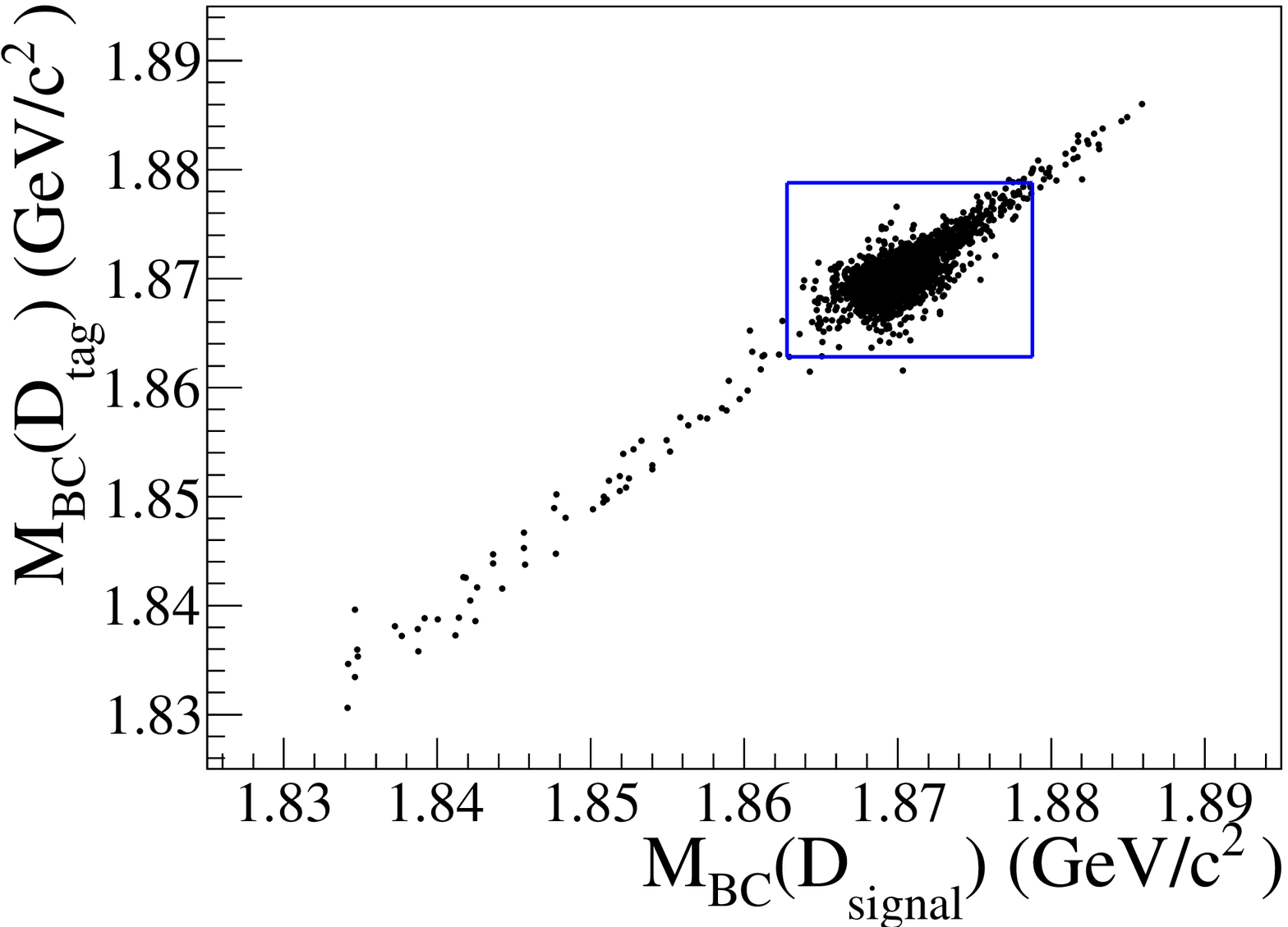}
\put(-130,100){(c)}
\end{minipage}
\begin{minipage}[b]{0.35\textwidth}
\epsfig{width=0.98\textwidth,file=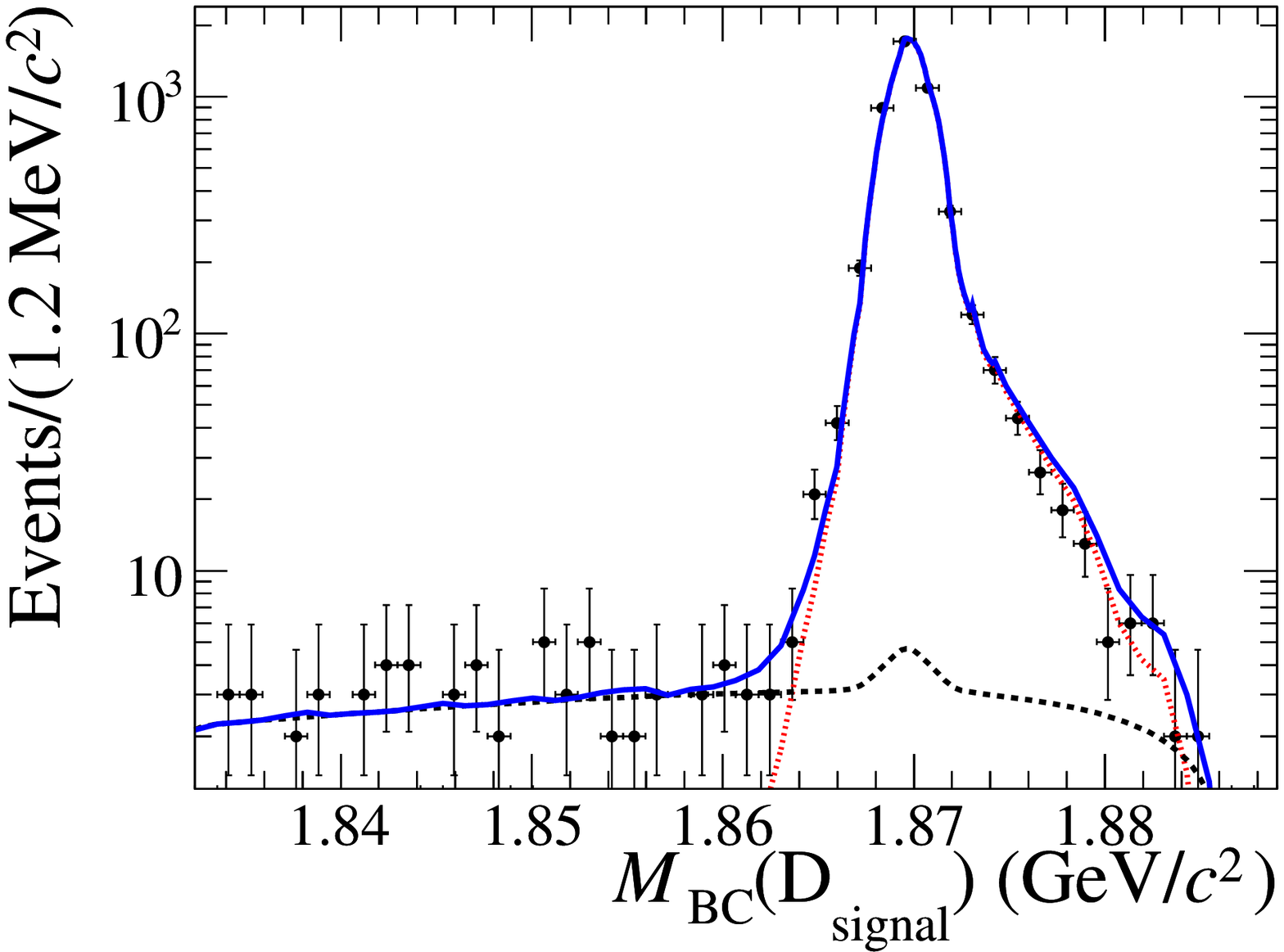}
\put(-130,100){(d)}
\end{minipage}
\caption{The $\Delta E$ distributions of data for (a) tag and (b) signal candidates, the (c) 2D distribution of 
$M_{{\rm BC}}(D_{{\rm tag}})$ versus $M_{{\rm BC}}(D_{{\rm signal}})$, 
and the (d) $M_{{\rm BC}}$ distribution for signal candidates. 
In (a), (b), and (d), the points with error bars are data. 
The arrows in (a) and (b) indicate the regions used to select the events, respectively. 
The rectangle in (c) shows the signal region. 
In (d), data are compared with the projection (solid curve) of the 2D fit, 
with the signal and the background marked as the dotted and dashed curves, respectively. 
The small bump under the signal is from the events with signal candidates properly reconstructed but tag candidates improperly reconstructed.}
\label{fig:selection}
\end{center}
\end{figure*}

The $D^{+}D^{-}$ pair with $D^{+} \rightarrow K_{S}^{0} \pi^{+} \pi^{+} \pi^{-}$  and $D^{-}\rightarrow K^{+} \pi^{-} \pi^{-}$ is reconstructed with the requirement that  
they do not have any tracks in common. 
If there are multiple $D^{+}D^{-}$ candidates reconstructed in an event, the one with the average invariant
mass closest to the nominal $D^{\pm}$ mass~\cite{PDG} is selected.
To characterize the $D$ candidates, two variables, $M_{{\rm BC}}$ and $\Delta E$, defined as
\begin{eqnarray}
\begin{aligned}
 M_{{\rm BC}} = \sqrt{E_{{\rm beam}}^2 -  \vec p_{D}^2}\\
\end{aligned}
\end{eqnarray}
and
\begin{eqnarray}
\begin{aligned}
\Delta E = E_{D} - E_{{\rm beam}},
\end{aligned}
\end{eqnarray}
are calculated,
where ($E_{D}$, $\vec p_{D}$) is the reconstructed four-momentum of $D$ candidate, and 
$E_{{\rm beam}}$ is the calibrated beam energy.
The signal events form a peak around zero in the $\Delta E$ distribution
and around the charged $D$ mass in the $M_{{\rm BC}}$ distribution.
Figures~\ref{fig:selection}(a-c) show the $\Delta E(D_{{\rm tag}})$ and $\Delta E(D_{{\rm signal}})$ distributions, 
and the two-dimensional (2D) distribution of $M_{{\rm BC}}(D_{{\rm tag}})$ versus $M_{{\rm BC}}(D_{{\rm signal}})$ of the accepted candidates in data, 
respectively. Events are required to satisfy $-0.027 < \Delta E (D_{{\rm tag}}) < 0.025$~GeV, $-0.033 < \Delta E (D_{{\rm signal}}) < 0.030$~GeV, 
and $1.8628 < M_{{\rm BC}} < 1.8788$~GeV/$c^{2}$ for both tag and signal $D$ candidates.

In order to suppress the background of $D^{+}\rightarrow K_{S}^{0}K_{S}^{0}\pi^{+}$  with an additional $K_S^0 \to \pi^+\pi^-$,  which has the same final state as our signal decay, we perform a decay vertex constrained fit on any remaining $\pi^{+}\pi^{-}$ pair with invariant mass within $\pm30$ MeV$/c^2$  of the mass of the $K_{S}^{0}$. The events are removed if the obtained decay length greater than twice of its uncertainty, . 
After applying all selection criteria,   the expected yield from the background $D^{+}\rightarrow K_{S}^{0}K_{S}^{0}\pi^{+}$  is estimated to be $72.9\pm8.5$ by using the generic MC sample.  
In the amplitude analysis, it is subtracted by giving negative weights to the background events, 
as discussed in Sec.~\ref{sec:lnL}.
Self cross-feed events with mis-reconstructed signal decays are estimated from 
signal MC samples to be $\sim0.1$\%. This effect is considered as a systematic uncertainty.

To estimate the contribution from the general background,
a 2D unbinned maximum likelihood fit is performed 
to the $M_{{\rm BC}}(D_{{\rm tag}})$ versus $M_{{\rm BC}}(D_{\rm signal})$ distribution in Fig.~\ref{fig:selection}(c). The signal shape is modeled with the MC-simulated shape. 
The diagonal background band is described by an ARGUS 
function~\cite{ARGUS} multiplied by a Gaussian in the anti-diagonal axis. 
The background with only the tag candidate (signal candidate) properly reconstructed peaks at 
 the charged $D$ mass and spreads out on the other axis, which
is parameterized as the product of a MC-simulated shape
in $M_{{\rm BC}}(D_{{\rm tag}})$ ($M_{{\rm BC}}(D_{{\rm signal}})$) and an ARGUS function on the other axis. 
The number of background events within the signal region extracted from the fit
is $37.5\pm7.5$.  
The projection on $M_{{\rm BC}}(D_{{\rm signal}})$ from the 2D fit is shown in Fig.~\ref{fig:selection}(d).
The small background bump under the signal is from the events with the $D_{{\rm signal}}$ properly
reconstructed but the $D_{{\rm tag}}$ improperly reconstructed. In the amplitude analysis, the general background is ignored and its
effect is considered as a systematic uncertainty.

To improve the momentum resolution and ensure that all events fall within the phase space boundary,
the selected candidate events are further subjected to a six-constraint (6C) kinematic fit.  It constrains the total four-momentum of all final state particles to the initial four-momentum of the $e^+e^-$ system,  
the invariant mass of signal side $D^+  \rightarrow K_{S}^{0} \pi^{+} \pi^{+} \pi^{-}$ constrains to the $D^+$ nominal mass, and the 
$K_{S}^{0}$ invariant mass constrains to the $K_S^0$ nominal mass. We discard events with a $\chi^{2}$ of 6C kinematic fit larger than $100$.
After applying all selection criteria, 4559 candidate events are obtained with a purity of 97.5\%.

\section{Amplitude Analysis}
\label{Amplitude Analysis}
The goal of this analysis is to determine the intermediate components  
in the four-body  $D^+\to K^0_S\pi^+\pi^+\pi^{-}$ decay. 
The decay modes which may contribute to the $D^+\to K^0_S\pi^+\pi^+\pi^{-}$ decay are listed in Table~\ref{tab:spin}. 
The letters $S$, $D$ in square brackets refer to the relative angular momentum between the daughter particles. 
The amplitudes and the relative phases between the different decay modes are determined with a maximum likelihood fit.

\subsection{Likelihood function construction}
\label{sec:lnL}
The unbinned maximum likelihood fit is performed by minimizing the 
negative log-likelihood (${\rm NLL}$) of the observed events ($N_{{\rm data}}$)
and the MC-simulated background events ($N_{{\rm bkg}}$):
\begin{eqnarray}
\begin{aligned}
{\rm NLL} = -\left[\sum_{k}^{N_{{\rm data}}}\ln f_{S}(p_{j}^{k})
      + \sum_{k^{\prime}}^{N_{{\rm bkg}}}w_{k^{\prime}}^{{\rm bkg}}\ln f_{S}(p_{j}^{k^{\prime}})\right],
\end{aligned}
\end{eqnarray}
where the indices $k$ and $k^{\prime}$ refer to the $k^{{\rm th}}$ event of the data
sample and the $k^{\prime {\rm th}}$ background event, respectively. The index $j$ refers to the $j^{{\rm th}}$ particle in the final state, 
$f_{S}(p_{j})$ is the signal probability density function (PDF) in terms of the final
four-momentum $p_{j}$, and $w_{k^{\prime}}^{{\rm bkg}}$ is the weight of the $k^{\prime {\rm th}}$ background event.
The contribution from the background is subtracted by assigning a negative weight to the background events.

The signal PDF $f_{S}(p_{j})$ is given by
\begin{eqnarray}
\begin{aligned}
\label{signalPDF}
f_{S}(p_{j}) = \frac{\epsilon(p_{j})|M(p_{j})|^{2}R_{4}(p_{j})}
{\int \epsilon(p_{j})|M(p_{j})|^{2}R_{4}(p_{j})dp_{j}},
\end{aligned}
\end{eqnarray}
where $M(p_{j})$ is the total decay amplitude describing the dynamics of
the $D^{+}$ decays, $\epsilon(p_{j})$ is the detection efficiency parameterized in terms of the final four-momentum $p_{j}$.
$R_{4}(p_{j})dp_{j}$ is the standard element of four-body phase space, which is given by
\begin{eqnarray}
\begin{aligned}
R_{4}(p_{j})dp_{j} =
\delta^{4}\left(p_{D^{0}}-\sum^{4}_{j}p_{j}\right)
\,\prod^{4}_{j}\frac{d^{3}{\bf p}_{j}}{(2\pi)^{3}2E_{j}}.
\end{aligned}
\end{eqnarray}
The $\epsilon(p_{j})$ in the numerator of Eq.~(\ref{signalPDF}) is independent of
the fitted variables, leading to a constant term in minimizing the likelihood and can be ignored in the fit.
The normalization integral of Eq.~(\ref{signalPDF}) is performed with a MC technique, which  
is then given by
\begin{eqnarray}
\begin{aligned}
&\int \epsilon(p_{j})|M(p_{j})|^{2}R_{4}(p_{j})dp_{j}\\
&=\frac{1}{N_{{\rm MC}}}\sum_{k_{{\rm MC}}}^{N_{{\rm MC}}}
  \frac{|M(p_{j}^{k_{{\rm MC}}}))|^{2}}
  {|M^{{\rm gen}}(p_{j}^{k_{{\rm MC}}})|^{2}},
\end{aligned}
\end{eqnarray}
where $k_{{\rm MC}}$ is the index of the $k^{{\rm th}}$ event of the MC sample and
$N_{{\rm MC}}$ is the number of the selected MC events.
$M^{{\rm gen}}(p_{j})$ is the PDF function used to generate the MC sample for the integration.

This analysis uses an isobar model formulation in which 
the total decay amplitude $M(p_{j})$ is given by the coherent sum over all contributing amplitudes:
\begin{eqnarray}
\begin{aligned}
M(p_{j}) = \sum_{n} \rho_{n}e^{i\phi_{n}} A_{n}(p_{j}),
\end{aligned}
\end{eqnarray}
where $\rho_{n}$ and $\phi_{n}$ are the magnitude and phase of the $n^\text{th}$ amplitude, respectively.
The $n^\text{th}$ amplitude $A_{n}(p_{j})$ is given by
\begin{eqnarray}
\begin{aligned}
A_{n}(p_{j}) = P_{n}^{1}(m_{1})P_{n}^{2}(m_{2})S_{n}(p_{j})B_{n}^{1}(p_{j})B_{n}^{2}(p_{j})B_{n}^{D}(p_{j}),
\end{aligned}
\end{eqnarray}
where the indexes 1 and 2 correspond to the two intermediate resonances.
$S_{n}(p_{j})$ is the spin factor, $P_{n}^{\alpha}(m_{\alpha})$ and $B_{n}^{\alpha}(p_{j})$ ($\alpha = 1,2$)
are the propagator and the Blatt-Weisskopf barrier factor~\cite{Blatt}, respectively,
and $B_{n}^{D}(p_{j})$ is the Blatt-Weisskopf barrier factor of the $D^{+}$ decay.
The parameters $m_{1}$ and $m_{2}$ in the propagators are the invariant masses of the corresponding resonances.
For non-resonant contributions with orbital angular momentum between the daughters,
we set the propagator to unity. This means that the amplitude has negligible $m$ dependence.
Since the $D^+ \to K_S^0 \pi^+\pi^+\pi^-$ decay contains two identical $\pi^{+}$s in the final state,
$A_{n}(p_{j})$ is symmetrized by exchanging the two $\pi^{+}$s to take into account the Bose symmetry.

\subsubsection{Spin factor}

The spin factor $S_{n}(p_{j})$ is constructed with the covariant tensor formalism~\cite{Zou}. 
The amplitudes with angular momenta larger than two are not considered due to the limited phase space.
For a specific process $a \rightarrow bc$,  
the covariant tensors $\tilde{t}^{L}_{\mu_{1}\cdot \cdot \cdot \mu_{l}}$ for the final states of pure orbital
angular momentum $L$ are constructed from the relevant momenta $p_{a}$, $p_{b}$, $p_{c}$~\cite{Zou}
\begin{eqnarray}
\begin{aligned}
\tilde{t}^{L}_{\mu_{1}\cdot \cdot \cdot \mu_{L}} =
(-1)^{L}P^{(L)}_{\mu_{1} \cdot \cdot \cdot \mu_{L}\nu_{1} \cdot \cdot \cdot \nu_{L}}
r^{\nu_{1}}\cdot \cdot \cdot r^{\nu_{L}},
\end{aligned}
\end{eqnarray}
where $r = p_{b} - p_{c}$.  
$P^{(L)}_{\mu_{1} \cdot \cdot \cdot \mu_{L}\nu_{1} \cdot \cdot \cdot \nu_{L}}$ 
is the spin projection operator and is defined as
\begin{eqnarray}
\begin{aligned}
P^{(1)}_{\mu \nu} = -g_{\mu\nu} + \frac{p_{a\mu}p_{a\nu}}{p^{2}_{a}}
\end{aligned}
\end{eqnarray}
for spin 1, and
\begin{eqnarray}
\begin{aligned}
&P^{(2)}_{\mu_{1}\mu_{2}\nu_{1}\nu_{2}} = \\
&\frac{1}{2}(P^{(1)}_{\mu_{1}\nu_{1}}P^{(1)}_{\mu_{2}\nu_{2}}+
P^{(1)}_{\mu_{1}\nu_{2}}P^{(1)}_{\mu_{2}\nu_{1}})
-\frac{1}{3}P^{(1)}_{\mu_{1}\mu_{2}}P^{(1)}_{\nu_{1}\nu_{2}}
\end{aligned}
\end{eqnarray}
for spin 2.

The spin factors of the decay modes used in the analysis are listed in Table~\ref{tab:spin}.
We use $\tilde{T}^{(L)}_{\mu_{1}...\mu_{L}}$ to represent the
decay of the $D^{+}$ meson and $\tilde{t}^{(L)}_{\mu_{1}...\mu_{L}}$
to represent the decay of the intermediate state.

\begin{table}[hbtp]
\footnotesize
\begin{center}
\caption{Spin factors $S(p)$ for different decay modes.}
\begin{tabular}{lc} \hline
Decay mode & $S(p)$ \\ \hline
$D\rightarrow \mbox{\,AP}_{1}, \mbox{\,A}[S]\rightarrow \mbox{\,VP}_{2}$,
$\mbox{\,V} \rightarrow \mbox{\,P}_{3} \mbox{\,P}_{4}$
&$\tilde{T}_{1}^{\mu}(D)P_{\mu\nu}^{(1)}(\mbox{\,A})\tilde{t}^{(1)\nu}(\mbox{\,V})$\\
$D\rightarrow \mbox{\,AP}_{1}, \mbox{\,A}[D]\rightarrow \mbox{\,VP}_{2}$,
$\mbox{\,V} \rightarrow \mbox{\,P}_{3} \mbox{\,P}_{4}$
& $\tilde{T}^{(1)\mu}(D)\tilde{t}_{\mu\nu}^{(2)}(\mbox{\,A})\tilde{t}^{(1)\nu}(\mbox{\,V})$\\
$D \rightarrow \mbox{\,AP}_{1}, \mbox{\,A} \rightarrow \mbox{\,SP}_{2} $,
$\mbox{\,S} \rightarrow \mbox{\,P}_{3} \mbox{\,P}_{4}$
& $\tilde{T}^{(1)\mu}(D)\tilde{t}^{(1)}_{\mu}(\mbox{\,A})$\\
$D \rightarrow \mbox{\,V}_{1}\mbox{\,P}_{1}, \mbox{\,V}_{1} \rightarrow \mbox{\,V}_{2}\mbox{\,P}_{2}$,
$\mbox{\,V}_{2} \rightarrow \mbox{\,P}_{3} \mbox{\,P}_{4}$
&$\epsilon_{\mu\nu\lambda\sigma}p_{\mbox{\,V}_{1}}^{\mu}q_{\mbox{\,V}_{1}}^{\nu}p_{\mbox{\,P}_{1}}^{\lambda}q_{\mbox{\,V}_{2}}^{\sigma}$\\
$D \rightarrow \mbox{\,PP}_{1}, \mbox{\,P} \rightarrow \mbox{\,VP}_{2}$,
$\mbox{\,V} \rightarrow \mbox{\,P}_{3} \mbox{\,P}_{4}$
& $p^{\mu}(\mbox{\,P}_{2})\tilde{t}^{(1)}_{\mu}(\mbox{\,V}) $\\
\hline
\end{tabular}
\label{tab:spin}
\end{center}
\end{table}

\subsubsection{Blatt-Weisskopf barrier factors}

For the process $a \rightarrow bc$,
the Blatt-Weisskopf barrier factor~\cite{Blatt} $B_{L}(p_{j})$ is parameterized as 
a function of the angular momentum $L$ and the momentum $q$ of the daughter $b$ or $c$ 
in the rest system of the $a$, 
\begin{eqnarray}
\begin{aligned}
B_{L}(q) = z^{L}X_{L}(q),
\end{aligned}
\end{eqnarray}
where $z = qR$. $R$ is the effective radius of the barrier, which is fixed to
$3.0{\mbox{\,GeV}^{-1}}$ for the intermediate resonances and
$5.0{\mbox{\,GeV}^{-1}}$ for the $D^{+}$ meson.
$X_{L}(q)$ is given by
\begin{eqnarray}
\begin{aligned}
X_{L = 0}(q) = 1,
\end{aligned}
\end{eqnarray}
\begin{eqnarray}
\begin{aligned}
X_{L = 1}(q) = \sqrt{\frac{2}{z^{2}+1}},
\end{aligned}
\end{eqnarray}
\begin{eqnarray}
\begin{aligned}
X_{L = 2}(q) = \sqrt{\frac{13}{z^{4}+3z^{2}+9}}.
\end{aligned}
\end{eqnarray}
With the invariant mass squared $s_{a/b/c}$ of the particle $ a/b/c$, the $q$ is 
\begin{eqnarray}
\begin{aligned}
\label{Qabc}
q = \sqrt{\frac{(s_{a}+s_{b}-s_{c})^2}{4s_{a}}-s_{b}}. 
\end{aligned}
\end{eqnarray}

\subsubsection{Resonance line shapes}
The propagator $P(m)$ describes the line shape of the intermediate resonance.
The resonances $\omega$, $K^{*-}$, $\bar{K}_{1}(1400)^{0}$, $a_{1}(1260)^{+}$ and $\bar{K}(1460)^{0}$ 
are parameterized
with a relativistic Breit-Wigner (RBW) line shape 
\begin{eqnarray}
\begin{aligned}
P^{{\rm RBW}}(m) = \frac{1}{(m^{2}_{0} - m^{2})-im_{0}\Gamma (m)},
\end{aligned}
\end{eqnarray}
where $m_{0}$ is the  mass of resonance  and $\Gamma (m)$ is the
mass-dependent width. The latter is expressed as
\begin{eqnarray}
\begin{aligned}
\label{Gamma_m}
\Gamma (m) = \Gamma_{0}\left(\frac{q}{q_{0}}\right)^{2L+1}\left(\frac{m_{0}}{m}\right)\left(\frac{X_{L}(q)}{X_{L}(q_{0})}\right)^{2},
\end{aligned}
\end{eqnarray}
where $\Gamma_0$ is the  width of resonance and $q_{0}$ denotes the value of $q$ at $m=m_{0}$.
The $\omega$ and $K_{1}(1270)^{-}$ are parameterized as a RBW with a constant width $\Gamma (m) = \Gamma_{0}$.

The resonance $\rho^{0}$ is described by the Gounaris-Sakurai (GS) function $P^{{\rm GS}}_{\rho}(m)$
with the $\rho-\omega$ interference taken into account~\cite{GS,LHCb}:
\begin{eqnarray}
\begin{aligned}
P_{\rho-\omega}(m) = P^{{\rm GS}}_{\rho}(m)(1 + \rho_{\omega}e^{i\phi_{\omega}}P^{{\rm RBW}}_{\omega}(m)),
\end{aligned}
\end{eqnarray}
where $\rho_{\omega}$ and $\phi_{\omega}$ are the relative magnitude and phase, respectively.
$P^{{\rm GS}}_{\rho}(m)$ is given by
\begin{eqnarray}
\begin{aligned}
P^{{\text GS}}_{\rho}(m) = \frac{1+d\frac{\Gamma_{0}}{m_{0}}}{(m_{0}^{2}-m^{2})+f(m)-im_{0}\Gamma (m)},
\end{aligned}
\end{eqnarray}
where
\begin{eqnarray}
\begin{aligned}
f(m) = \Gamma_{0}\frac{m_{0}^{2}}{q_{0}^{3}}[q^{2}(h(m)-h(m_{0}))\\
+(m_{0}^{2}-m^{2})q_{0}^{2}\frac{dh}{d(m^{2})}\mid_{m^{2} = m^{2}_{0}}],&
\end{aligned}
\end{eqnarray}
and the function $h(m)$ is defined as
\begin{eqnarray}
\begin{aligned}
h(m) = \frac{2}{\pi}\frac{q}{m}\ln(\frac{m+2q}{2m_{\pi}}),
\end{aligned}
\end{eqnarray}
with
\begin{eqnarray}
\begin{aligned}
&\frac{dh}{d(m^{2})}\Big|_{m^{2} = m^{2}_{0}} = \\
&h(m_{0})[(8q_{0}^{2})^{-1}-(2m_{0}^{2})^{-1}]+(2\pi m_{0}^{2})^{-1},
\end{aligned}
\end{eqnarray}
where $m_{\pi}$ is the charged pion mass~\cite{PDG}.
The normalization condition at $P^{\text{GS}}(0)$ fixes the parameter $d=f(0)/(\Gamma_{0}m_{0})$.
It is found to be~\cite{GS}
\begin{eqnarray}
\begin{aligned}
d = \frac{3}{\pi}\frac{m_{\pi}^{2}}{q_{0}^{2}}\ln\left(\frac{m_{0}+2q_{0}}{2m_{\pi}}\right)+\frac{m_{0}}{2\pi q_{0}}
- \frac{m_{\pi}^{2}m_{0}}{\pi q_{0}^{3}}.
\end{aligned}
\end{eqnarray}

The resonance $f_{0}(500)$ is parameterized with the formula given in Ref.~\cite{Bugg:2003kj}:
\begin{eqnarray}
\begin{aligned}
P_{f_{0}(500)}(m)= \frac{1}{m_{0}^{2}-m^{2}-im_{0}\Gamma_{{\rm tot}}(m)},
\end{aligned}
\end{eqnarray}
where $\Gamma_{{\rm tot}}(m)$ is decomposed into two parts: 
\begin{eqnarray}
\begin{aligned}
\Gamma_{{\rm tot}}(m) = g_{1}\frac{\rho_{\pi\pi}(m)}{\rho_{\pi\pi}(m_{0})} + g_{2}\frac{\rho_{4\pi}(m)}{\rho_{4\pi}(m_{0})} 
\end{aligned}
\end{eqnarray}
and 
\begin{eqnarray}
\begin{aligned}
g_{1} = (b_{1}+b_{2}m^{2})\frac{m^{2}-m_{\pi}^{2}/2}{m^{2}_{0}-m_{\pi}^{2}/2}e^{(m_{0}^{2}-m^{2})/a}.
\end{aligned}
\end{eqnarray}
Here is $\rho_{\pi\pi}$ the phase space of the $\pi^{+}\pi^{-}$ 
system and $\rho_{4\pi}$ is the $4\pi$ phase space approximated by 
$\sqrt{1-16m^{2}_{\pi}/m^{2}}/[1+e^{(2.8-m^{2})/3.5}]$. 
The parameters are fixed to the values given in Ref.~\cite{Ablikim:2004qna}. 

The resonance $K^{*}(1430)^{-}$ is considered in a $K\pi$ $S$-wave 
(denoted as $(K_{S}^{0}\pi^{-})_{S{\rm -wave}}$) parameterization
extracted from the scattering data~\cite{LASS}. 
The same parameterization was used by {\sc BABAR} in Ref.~\cite{KPiS}:
\begin{eqnarray}
\begin{aligned}
P^{S-{\rm wave}}(m_{K\pi}) = F\sin\delta_{F}e^{i\delta_{F}} + R\sin\delta_{R}e^{i\delta_{R}}e^{i2\delta_{F}},
\end{aligned}
\end{eqnarray}
with
\begin{align}
\delta_{F} &= \phi_{F} +
             \cot^{-1}\left[\frac{1}{aq}+\frac{rq}{2}\right],\\
\delta_{R} &= \phi_{R} + \tan^{-1}\left[\frac{M\Gamma(m_{K\pi})}{M^{2}-m_{K\pi}^{2}}\right],
\end{align}
where $a$ and $r$ denote the scattering length and effective interaction length, respectively.
$F(\phi_{F})$ and $R(\phi_{R})$ are the relative magnitudes (phases) for
the non-resonant and resonant terms, and
$q$ and $\Gamma(m_{K\pi})$ are defined as in Eq.~(\ref{Qabc}) and Eq.~(\ref{Gamma_m}), respectively.
In the fit, the parameters $M$, $\Gamma$, $F$, $\phi_{F}$, $R$, $\phi_{R}$, $a$ and $r$
are fixed to the values obtained from the fit to the
$D^{0} \rightarrow K_{S}^{0}\pi^{+}\pi^{-}$ Dalitz plot performed by {\sc BABAR}~\cite{KPiS}
and are given in Table ~\ref{tab:BABAR KPiS}.
\begin{table}[htp]
\begin{center}
\caption{$(K_{S}^{0}\pi^{-})_{S{\rm -wave}}$  parameters,
obtained from the fit to the $D^{0} \rightarrow K_{S}^{0}\pi^{+}\pi^{-}$
Dalitz plot from BABAR~\cite{KPiS}. The uncertainties are statistical. }
\begin{tabular}{cc} \hline
$M$(GeV/$c^{2}$) & $1.463\pm0.002$ \\
$\Gamma$(GeV/$c^{2}$) & $0.233\pm0.005$ \\
$F$ & $0.80\pm0.09$ \\
$\phi_{F}$ & $2.33\pm0.13$ \\
$R$ & $1$(fixed)\\
$\phi_{R}$ & $-5.31\pm0.04$ \\
$a$ & $1.07\pm0.11$ \\
$r$ & $-1.8\pm0.3$ \\
\hline
\end{tabular}
\label{tab:BABAR KPiS}
\end{center}
\end{table}

\subsection{Fit fraction}
\label{FF_cal}
The fit fraction (FF) for an amplitude or a component (a certain subset of amplitudes) is
calculated using a large set of generation-level PHSP MC sample by 
\begin{eqnarray}
\begin{aligned}
\label{eq_FF}
{\rm FF}(n)  = \frac{\sum_{k=1}^{N_{{\rm gen}}}|\tilde{A}_{\bf n}(p_{j}^{k})|^{2}}
{\sum_{k=1}^{N_{{\rm gen}}}|M(p_{j}^{k})|^{2}},
\end{aligned}
\end{eqnarray}
where $\tilde{A}_{\bf n}(p_{j}^{k})$ is either the $n^{{\rm th}}$ amplitude
($\tilde{A}_{{\bf n}}(p_{j}^{k}) = \rho_{n}e^{i\phi_{n}}A_{n}(p_{j}^{k})$)
or the ${\bf n}^{{\rm th}}$ component of a coherent sum of amplitudes
($\tilde{A}_{{\bf n}}(p_{j}^{k}) = \sum{\rho_{n_{i}}e^{i\phi_{n_{i}}}A_{n_{i}}(p_{j}^{k})}$) and
$N_{{\rm gen}}$ is the number of the PHSP MC events.
Note that the sum of the FFs is not necessarily equal to unity due to the
interferences among the contributing amplitudes.

To obtain the statistical uncertainties of the FFs, 
the FFs are calculated 500 times by randomly varying the floated parameters according to the full covariance matrix. 
The distribution for each amplitude or each component 
is fitted with a Gaussian function. 
The width of the Gaussian function is the statistical uncertainty of the corresponding FF.
 
\subsection{Goodness of fit}
\label{Goodness of Fit}
To examine the performance of the fit process, the goodness of fit is defined as follows.
Since the $D^{+}$ and all four final-state particles have spin zero, the phase space of the decay
can be completely described by five linearly independent Lorentz invariant variables.
For convenience, one of the two identical  pions which results in a lower
$\pi^{+}\pi^{-}$ invariant mass is denoted as $\pi^{+}_{1}$, while the other as $\pi_{2}^{+}$.
The four final-state particles $K_{S}^{0}$, $\pi^{+}_{1}$, $\pi^{+}_{2}$, $\pi^{-}$ are marked with the indices 
$1,\,2,\,3,\,4,$ respectively.
Then the five invariant masses
$m_{24}$, $m_{34}$, $m_{124}$, $m_{134}$, and $m_{234}$ are chosen to map out the phase space.
To calculate the goodness of fit, the five-dimensional phase space is
first divided into cells with
equal size. Then, adjacent cells are combined until the number of events in each cell is larger than 20.
The fit residual $\chi$ in $p^{{\rm th}}$ cell is calculated, $\chi_{p} = \frac{N_{p} - N_{p}^{{\rm exp}}}{\sqrt{N_{p}^{{\rm exp}}}}$.
The goodness of fit is quantified as $\chi^{2} = \sum_{p=1}^{n}\chi_{p}^{2}$,
where $N_{p}$ and $N_{p}^{{\rm exp}}$ are the number of observed and
expected number of events from the fit in the $p^{{\rm th}}$ cell, respectively,
and $n$ is the total number of cells.
The number of degrees of freedom (NDF) $\nu$ is given by $\nu = (n-1) - n_{{\rm par}}$, where $n_{{\rm par}}$
is the number of the free parameters in the fit.

\section{Results}
\label{RESULTS}
\label{Nominal Fit}

We start the fit of the data by considering the amplitudes 
containing ${K}^{*-}$, $\rho^{0}$, $\bar{K}_{1}(1270)^{0}$, $\bar{K}_{1}(1400)^{0}$,
$a_{1}(1260)^{+}$ resonances, as these resonances are clearly observed in the 
corresponding invariant mass spectra.
We then add amplitudes with resonances listed in the PDG~\cite{PDG} and non-resonant components until no additional 
amplitude has a significance larger than $5\sigma$.
The statistical significance for any new amplitude is
calculated with the change of the log-likelihood value $\Delta ({\rm NLL})$ after taking the change of the degrees of freedom $\Delta \nu$ into account.
In the fits, the amplitude and phase of 
$D^{+}\rightarrow K_{S}^{0}a_{1}(1260)^{+}(\rho^{0}\pi^{+}[S])$ are fixed to 1 and 0 as the 
reference, while the magnitudes and phases of the other amplitudes are floating. 
Here, [S]  means the angular momentum of $\rho^{0}\pi^{+}$ combination is zero ($S$-wave).  
The corresponding $D$-wave amplitude $D^{+}\rightarrow K_{S}^{0}a_{1}(1260)^{+}(\rho^{0}\pi^{+}[D])$ 
is found to have a FF of about 1\% of the $S$-wave, which is consistent with
both BESIII and LHCb amplitude analyses on $D^{0} \rightarrow K^{-}\pi^{+}\pi^{+}\pi^{-}$~\cite{Ablikim:2017eqz,Aaij:2017kbo}. 
We consider therefore this $D$-wave amplitude in the nominal fit although its significance is 4.3$\sigma$.  

The resonant term $D^{+}\rightarrow K_{S}^{0}a_{1}(1260)^{+}(\rho^{0}\pi^{+}[S])$ and its non-resonant partner    
$D^{+}\rightarrow K_{S}^{0}(\rho^{0}\pi^{+}[S])_{A}$ (the subscript $A$ represents the 
axial-vector non-resonant state for the $\rho^{0}\pi^{+}$ combination), 
are both found with significances greater than 10$\sigma$, 
while they are highly correlated because of the same angular distribution and large common region in phase space.
For the resonant term in the fit model with the non-resonant partner, 
its FF becomes highly uncertain and is significantly different 
with the one in the fit model without the non-resonant partner. 
However the combined FF of these two amplitudes is almost unchanged. 
We, therefore, only consider the resonant term. 
Similar cases are also found with the amplitude pairs of $D^{+} \rightarrow \bar{K}(1460)^{0}(K_{S}^{0}\rho^{0})\pi^{+}$ 
and $D^{+}\rightarrow (K_{S}^{0}\rho^{0})_{P}\pi^{+}$, 
$D^{+} \rightarrow \bar{K}(1460)^{0}(K^{*-}\pi^{+})\pi^{+}$ and $D^{+}\rightarrow (K^{*-}\pi^{+})_{P}\pi^{+}$, 
as well as $D^{+} \rightarrow \bar{K}_{1}(1650)^{0}(K^{*-}\pi^{+}[S])\pi^{+}$ and $D^{+}\rightarrow (K^{*-}\pi^{+}[S])_{A}\pi^{+}$. 
Throughout this paper, we denote $K^{*-} \to K_{S}^{0} \pi^{-}$ and $\rho^{0} \to \pi^{+} \pi^{-}$, which is also included in the FFs and 
BFs of corresponding sub modes.
In the nominal fit, we only use the resonant terms, as done in the analysis of Mark III~\cite{MarKIII}. 

\begin{figure*}[hbtp]
\begin{center}
\begin{minipage}[b]{0.28\textwidth}
\epsfig{width=1.00\textwidth,clip=true,file=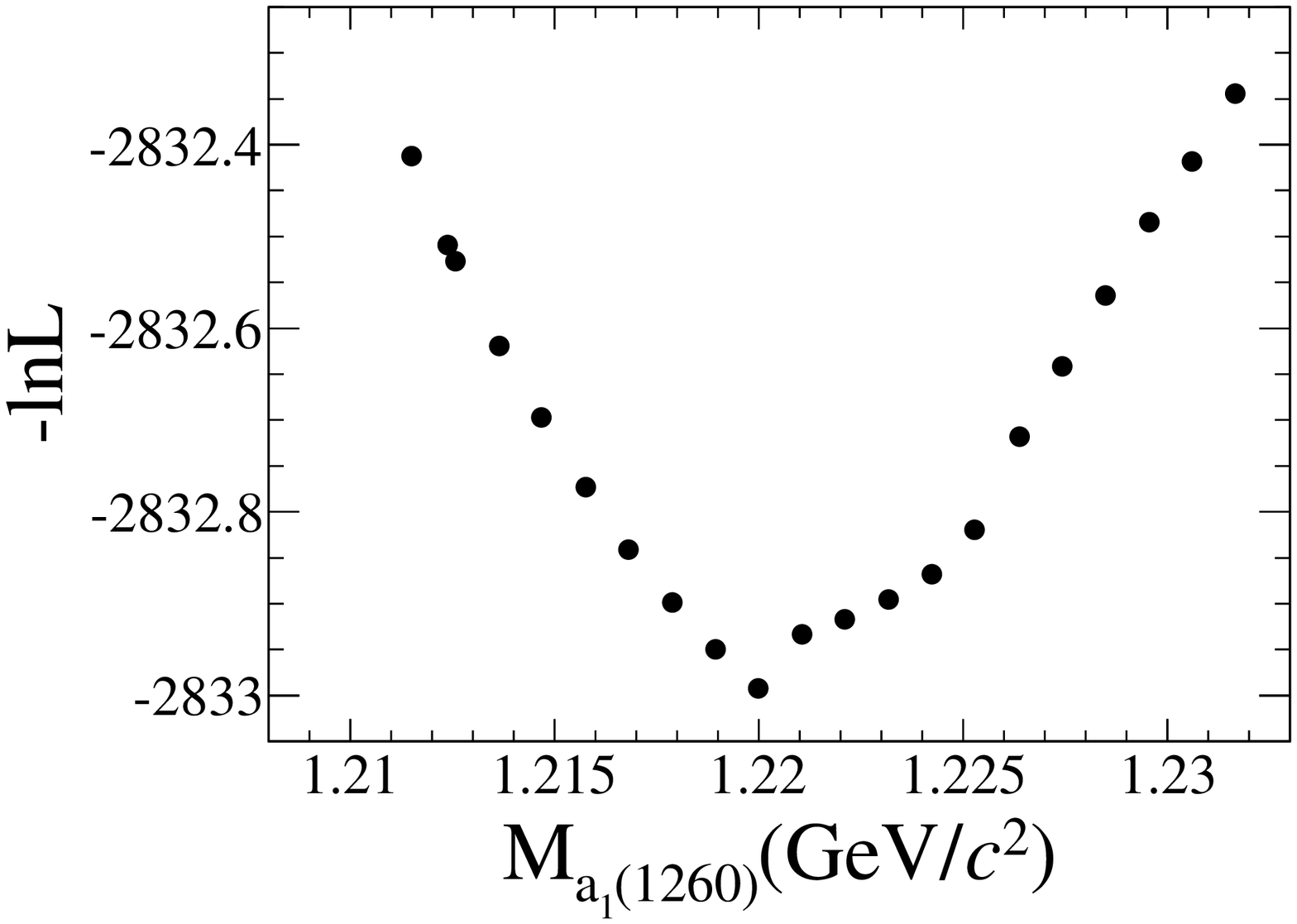}
\put(-70,75){(a)}
\end{minipage}
\begin{minipage}[b]{0.28\textwidth}
\epsfig{width=1.00\textwidth,clip=true,file=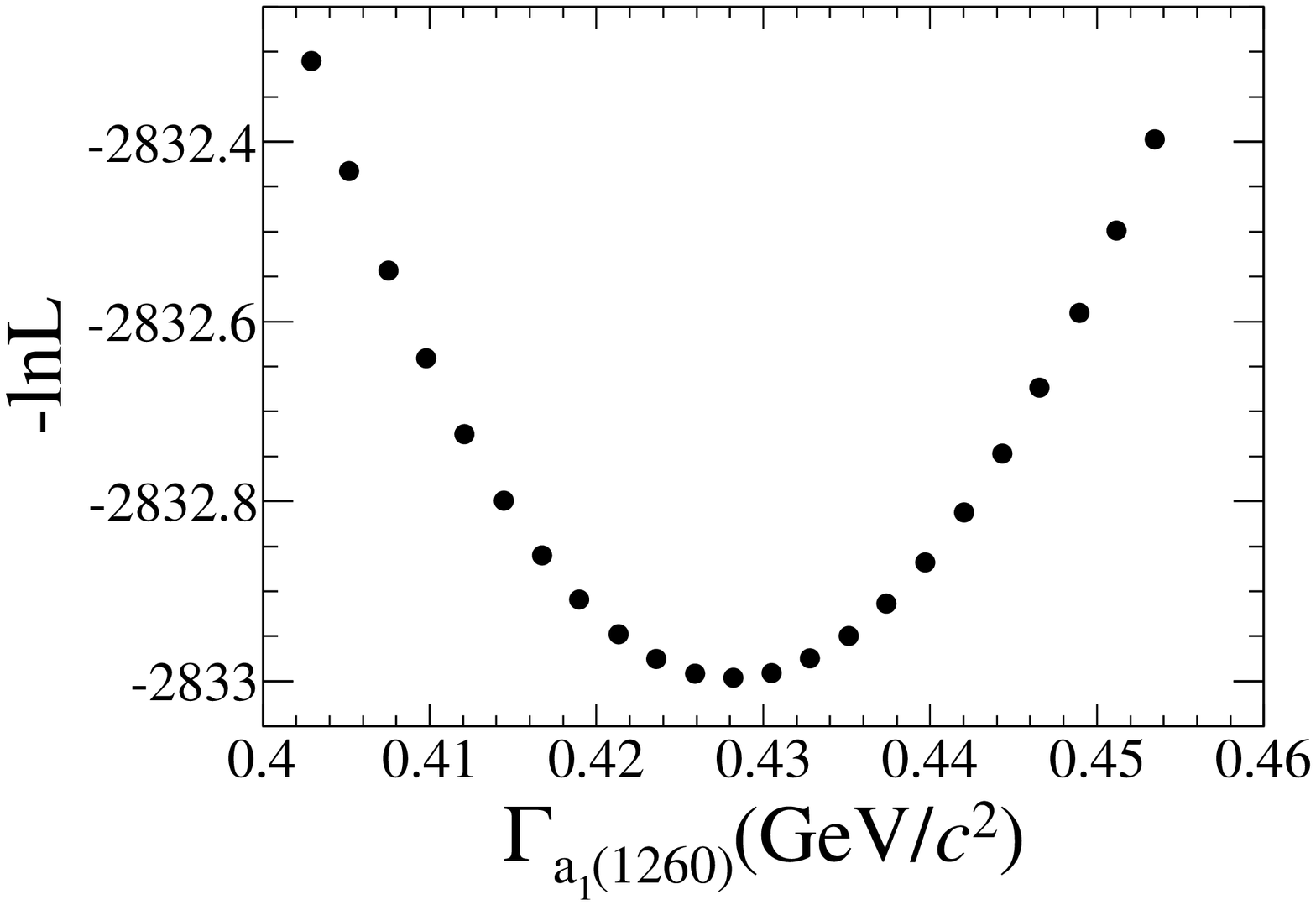}
\put(-70,75){(b)}
\end{minipage}
\begin{minipage}[b]{0.28\textwidth}
\epsfig{width=1.00\textwidth,clip=true,file=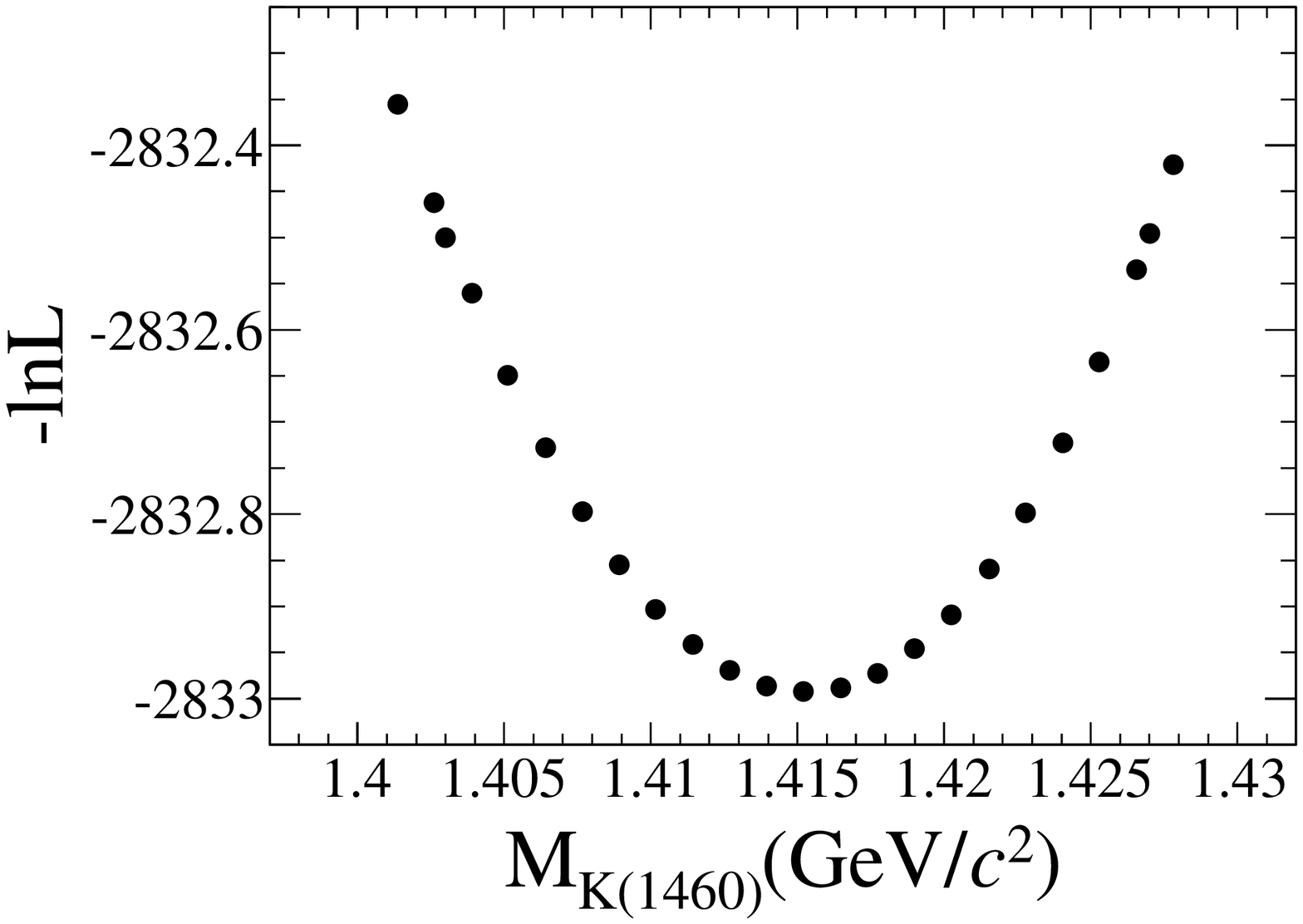}
\put(-70,75){(c)}
\end{minipage}
\begin{minipage}[b]{0.28\textwidth}
\epsfig{width=1.00\textwidth,clip=true,file=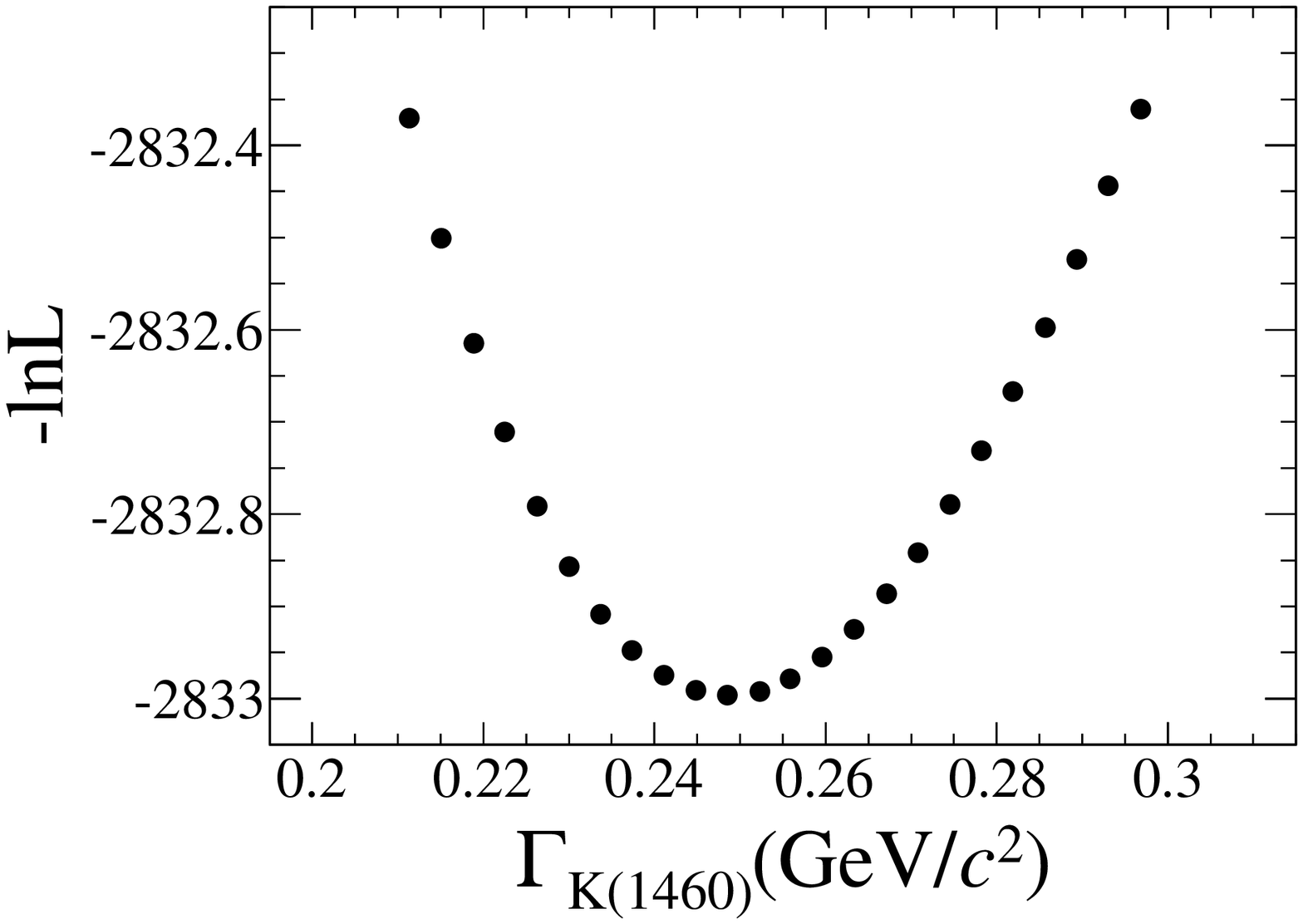}
\put(-70,75){(d)}
\end{minipage}
\begin{minipage}[b]{0.28\textwidth}
\epsfig{width=1.00\textwidth,clip=true,file=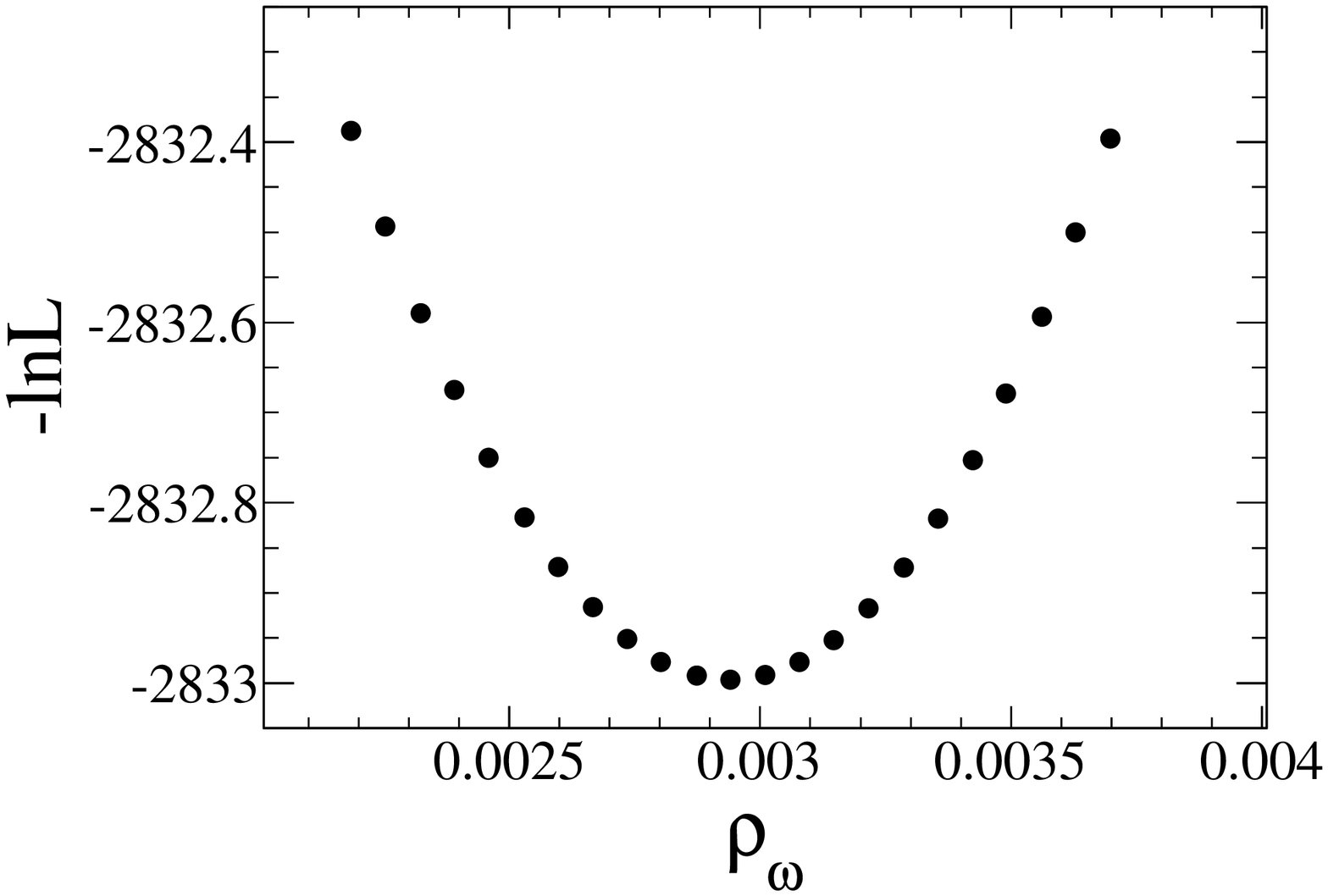}
\put(-70,75){(e)}
\end{minipage}
\begin{minipage}[b]{0.28\textwidth}
\epsfig{width=1.00\textwidth,clip=true,file=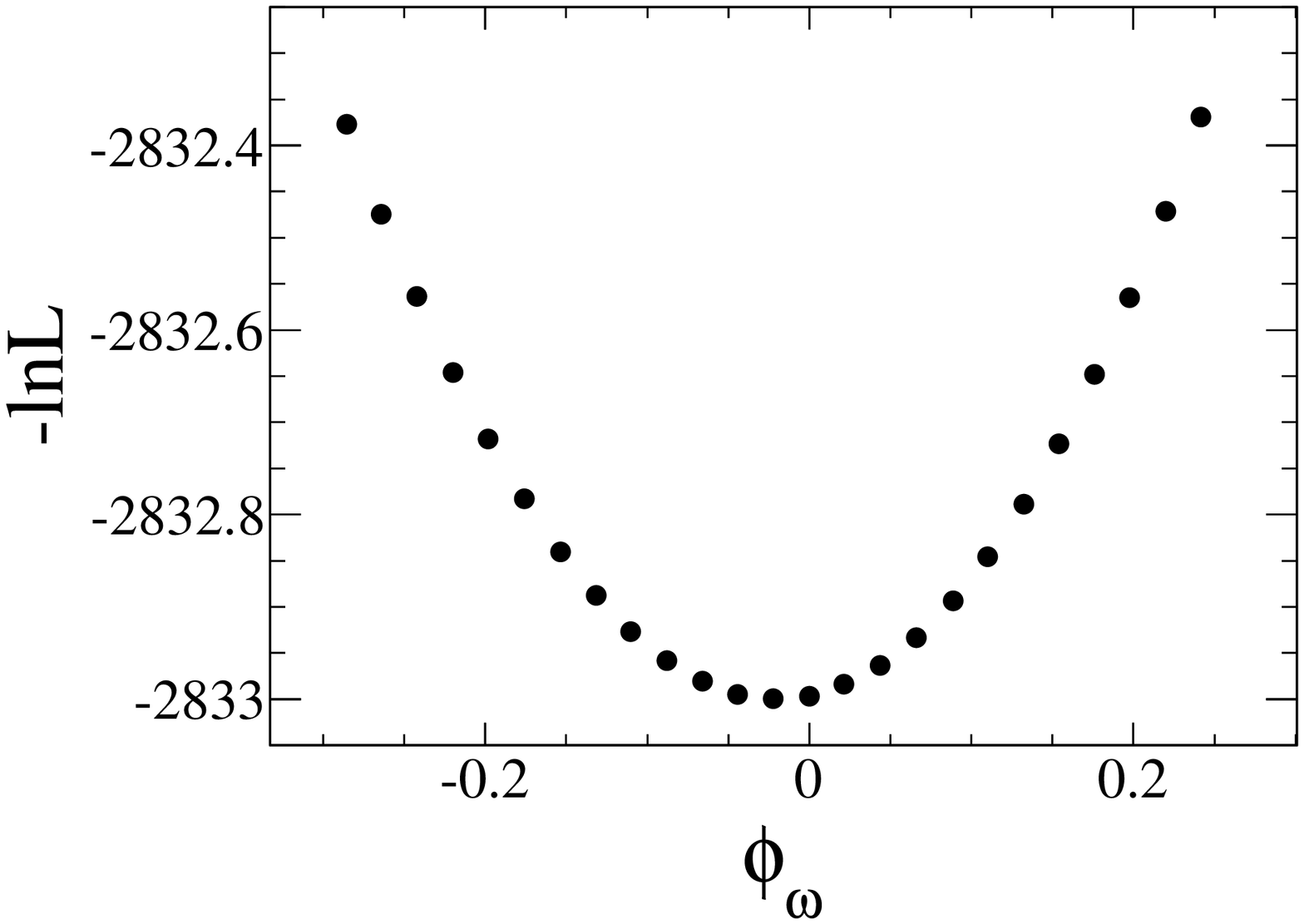}
\put(-70,75){(f)}
\end{minipage}
\caption{Likelihood scans of the masses ((a) and (c)) and widths
((b) and (d)) of $a_{1}(1260)^{+}$ and $\bar{K}(1460)^{0}$, respectively; as well as the
relative magnitude (e) and phase (f) of $\omega$ in the $\rho-\omega$ mixing.}
\label{fig:scan}
\end{center}
\end{figure*}

The masses and widths of $\rho^{0}$, $\omega$, ${K}^{*-}$, $\bar{K}_{1}(1270)^{0}$, 
$\bar{K}_{1}(1400)^{0}$, and $\bar{K}_{1}(1650)^{0}$ are
fixed at the values from  PDG~\cite{PDG}.
Since there are no world average values for the masses and widths of $a_{1}(1260)^{+}$ and $\bar{K}(1460)^{0}$ and the 
resonances lie on the upper boundary of the corresponding invariant mass spectrum, their values are determined 
by likelihood scans. 
The values of the parameters related to $\rho-\omega$ mixing are also determined by likelihood scans.
The scan results are: 
\begin{eqnarray}
\begin{aligned}
&m_{a_{1}(1260)^{+}} = 1220.0^{+9.5}_{-7.6} ~\mbox{\,MeV}/c^{2}, \\
&\Gamma_{a_{1}(1260)^{+}} = 428.2^{+23.0}_{-22.2} ~\mbox{\,MeV}/c^{2},\\
&m_{\bar{K}(1460)^{0}} = 1415.2^{+11.8}_{-12.2} ~\mbox{\,MeV}/c^{2}, \\
&\Gamma_{\bar{K}(1460)^{0}} = 248.5^{+40.8}_{-33.4} ~\mbox{\,MeV}/c^{2},\\
&\rho_{\omega} = (2.94\pm0.69)\times 10^{-3}, \\
&\phi_{\omega} = -0.02\pm0.23,
\end{aligned}
\end{eqnarray}
where the uncertainties are statistical only. In the nominal fit, these parameters are set to be the values 
determined by likelihood scans. The scan results are shown in Fig.~\ref{fig:scan}. In Fig.~\ref{fig:scan}(a), three scan points at the right of the minimum point 
are higher than smooth scan expectations due to the correlation between the states with resonances $a_{1}(1260)^{+}$ or $\bar{K}(1460)^{0}$ involved.

\begin{figure*}[hbtp]
\centering
\begin{minipage}[b]{0.28\textwidth}
\epsfig{width=1.00\textwidth,clip=true,file=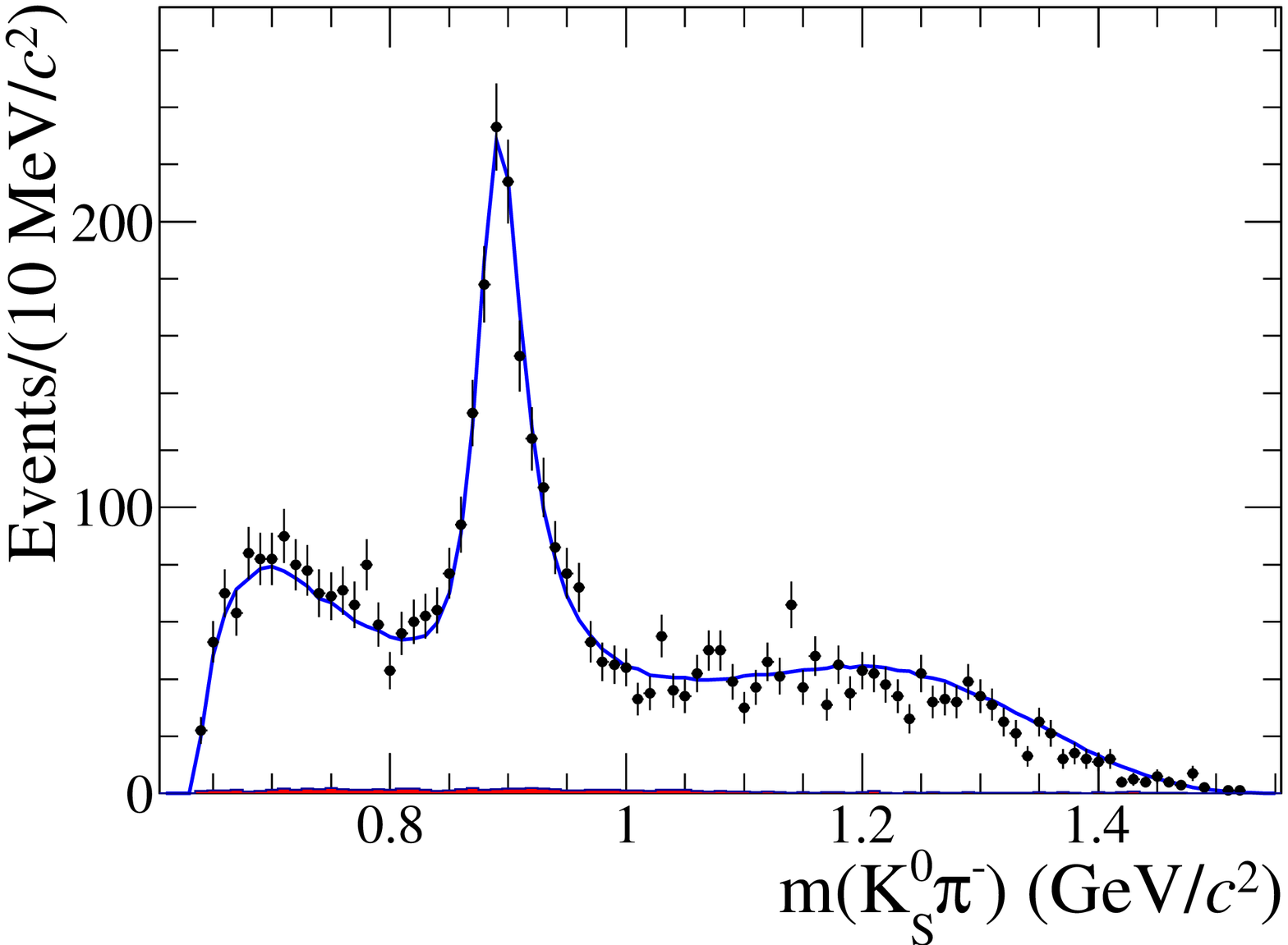}
\put(-25,90){(a)}
\end{minipage}
\begin{minipage}[b]{0.28\textwidth}
\epsfig{width=1.00\textwidth,clip=true,file=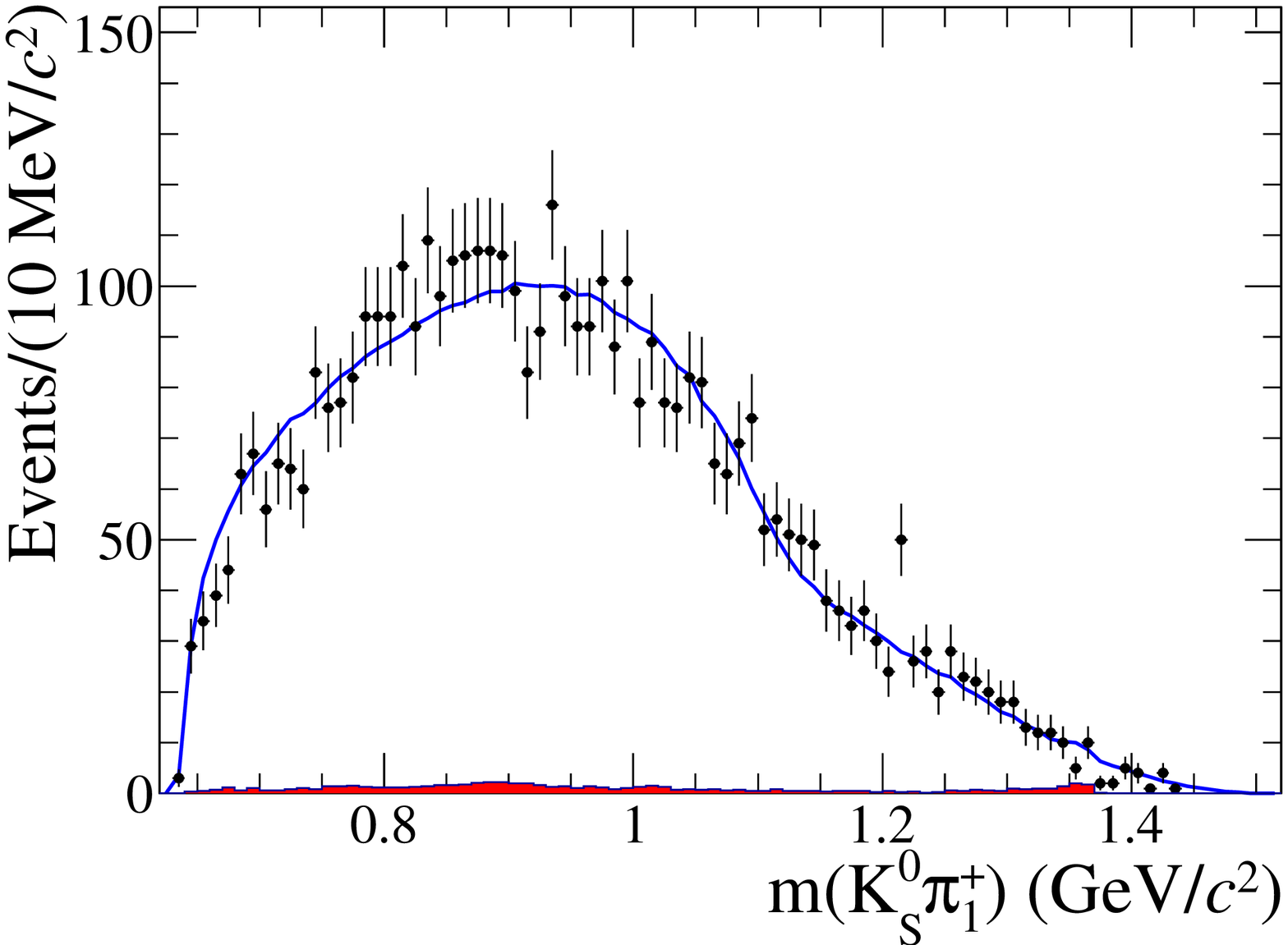}
\put(-25,90){(b)}
\end{minipage}
\begin{minipage}[b]{0.28\textwidth}
\epsfig{width=1.00\textwidth,clip=true,file=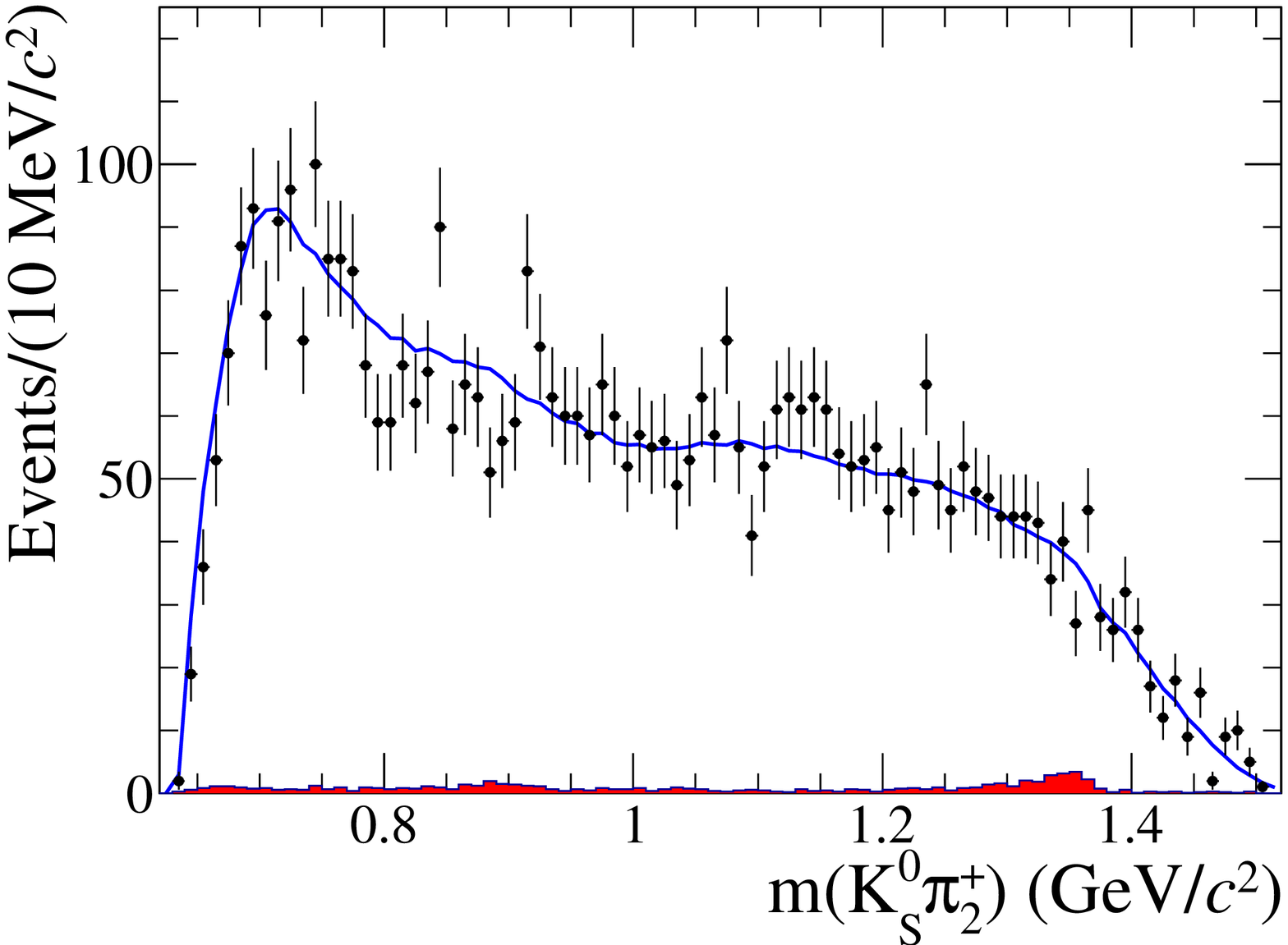}
\put(-25,90){(c)}
\end{minipage}
\begin{minipage}[b]{0.28\textwidth}
\epsfig{width=1.00\textwidth,clip=true,file=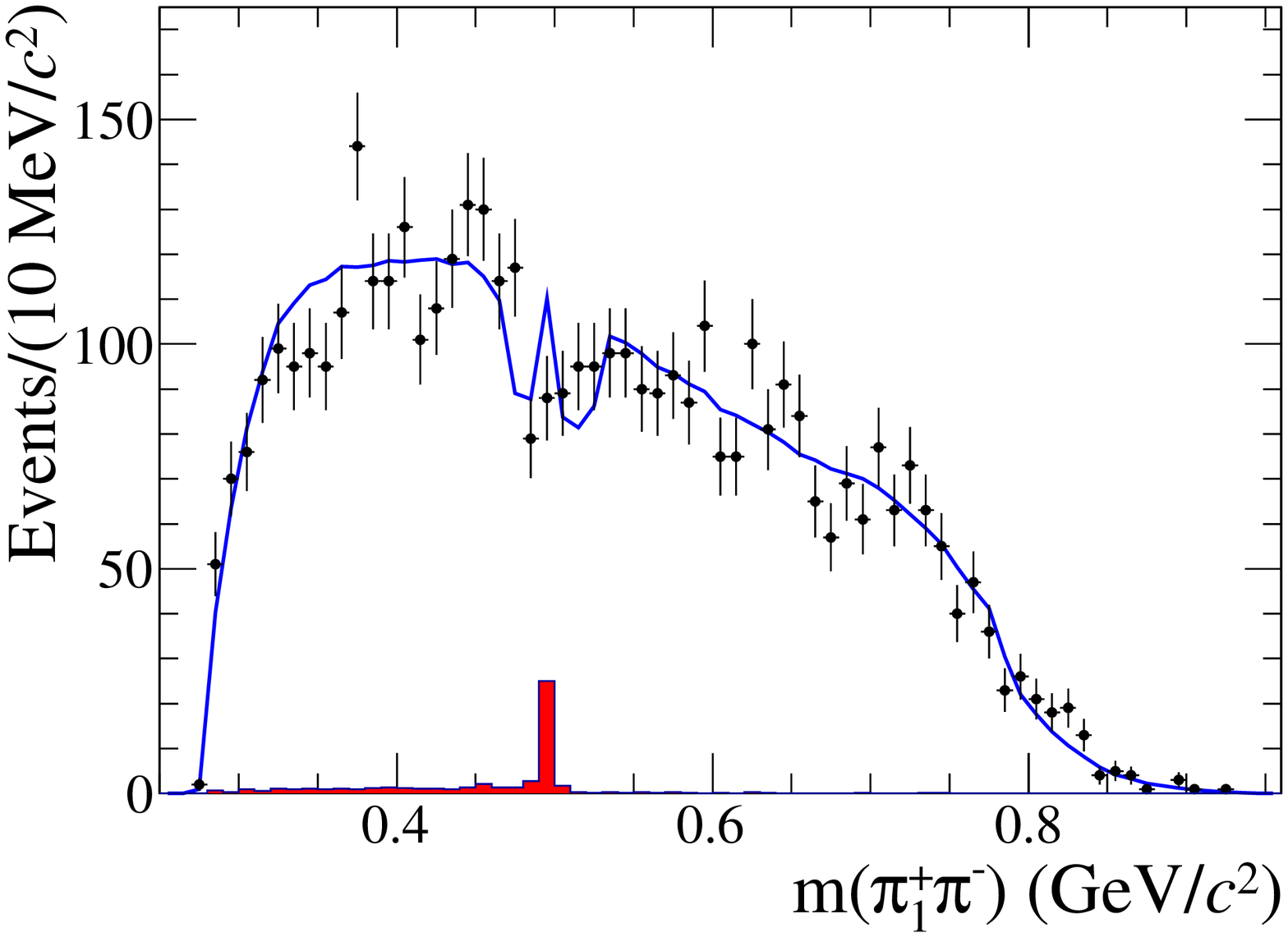}
\put(-25,90){(d)}
\end{minipage}
\begin{minipage}[b]{0.28\textwidth}
\epsfig{width=1.00\textwidth,clip=true,file=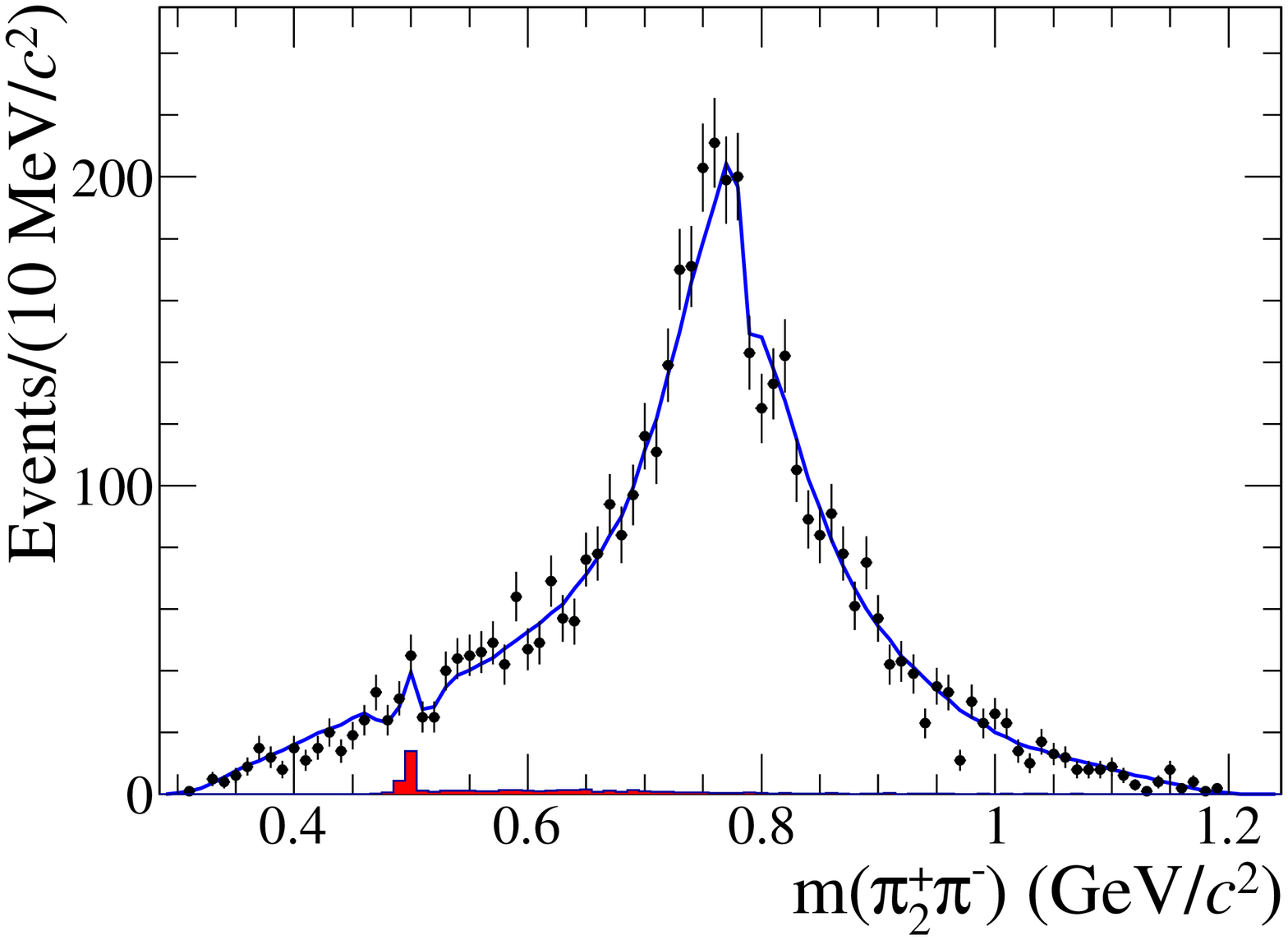}
\put(-25,90){(e)}
\end{minipage}
\begin{minipage}[b]{0.28\textwidth}
\epsfig{width=1.00\textwidth,clip=true,file=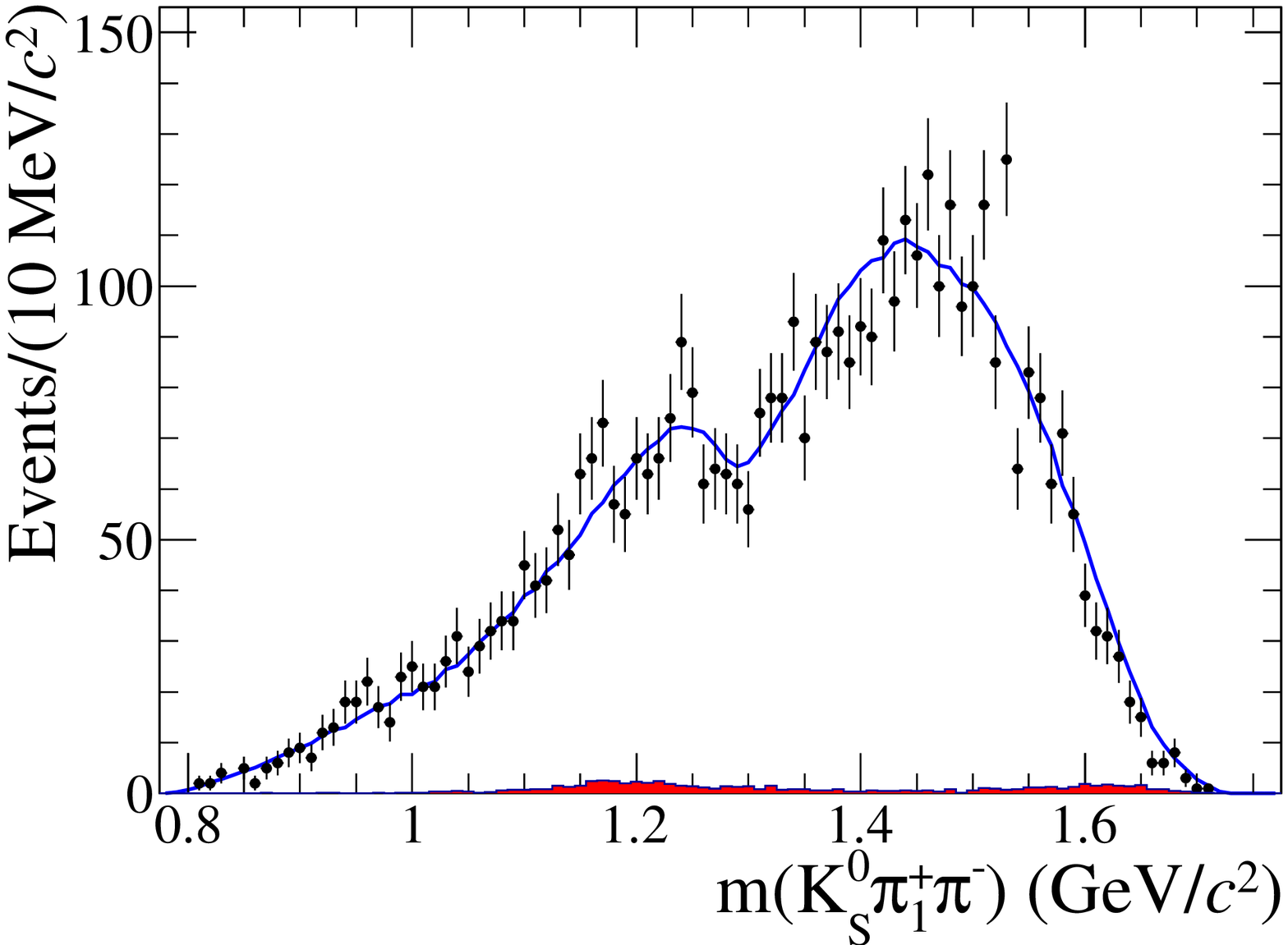}
\put(-25,90){(f)}
\end{minipage}
\begin{minipage}[b]{0.28\textwidth}
\epsfig{width=1.00\textwidth,clip=true,file=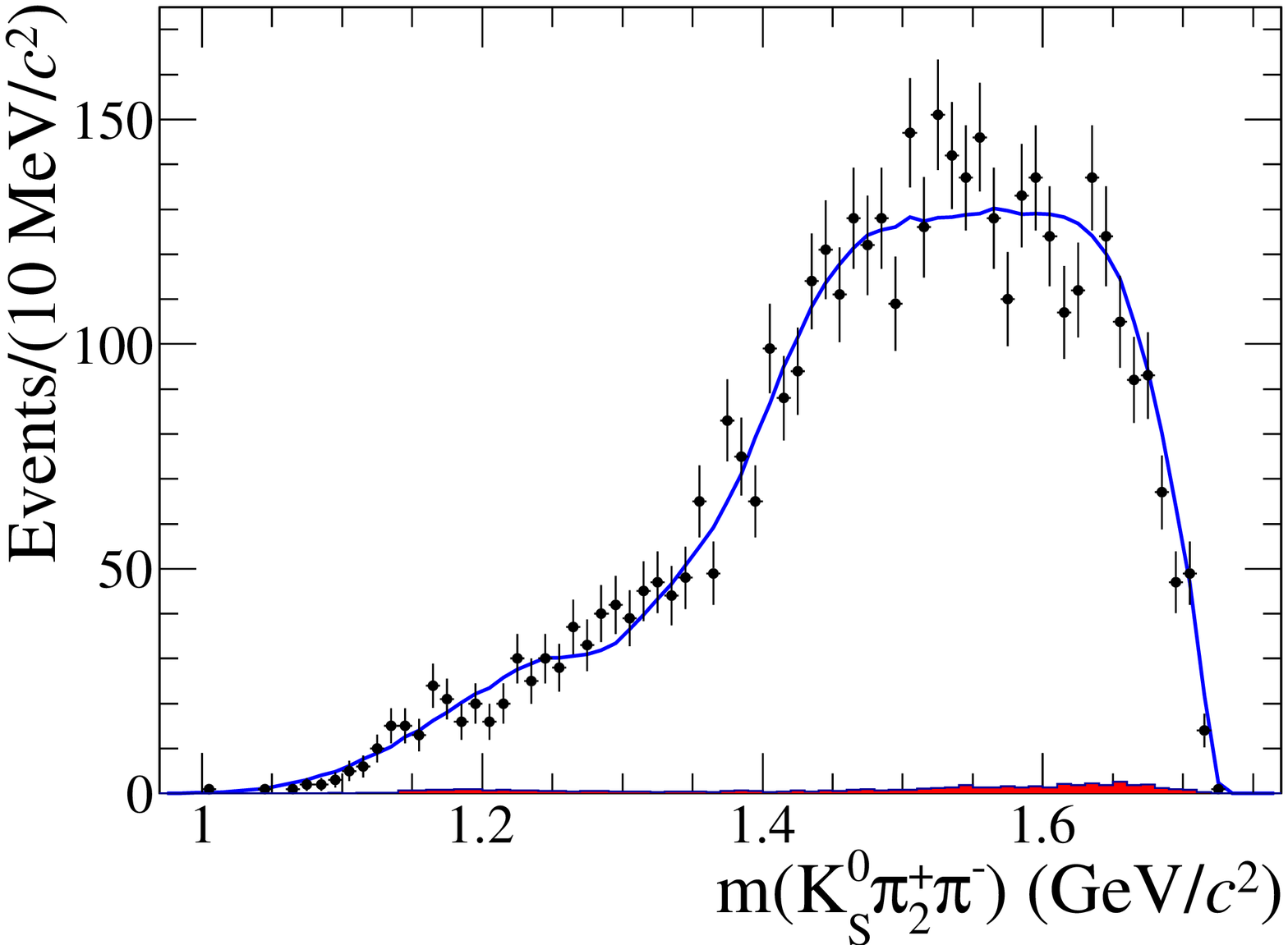}
\put(-25,90){(g)}
\end{minipage}
\begin{minipage}[b]{0.28\textwidth}
\epsfig{width=1.00\textwidth,clip=true,file=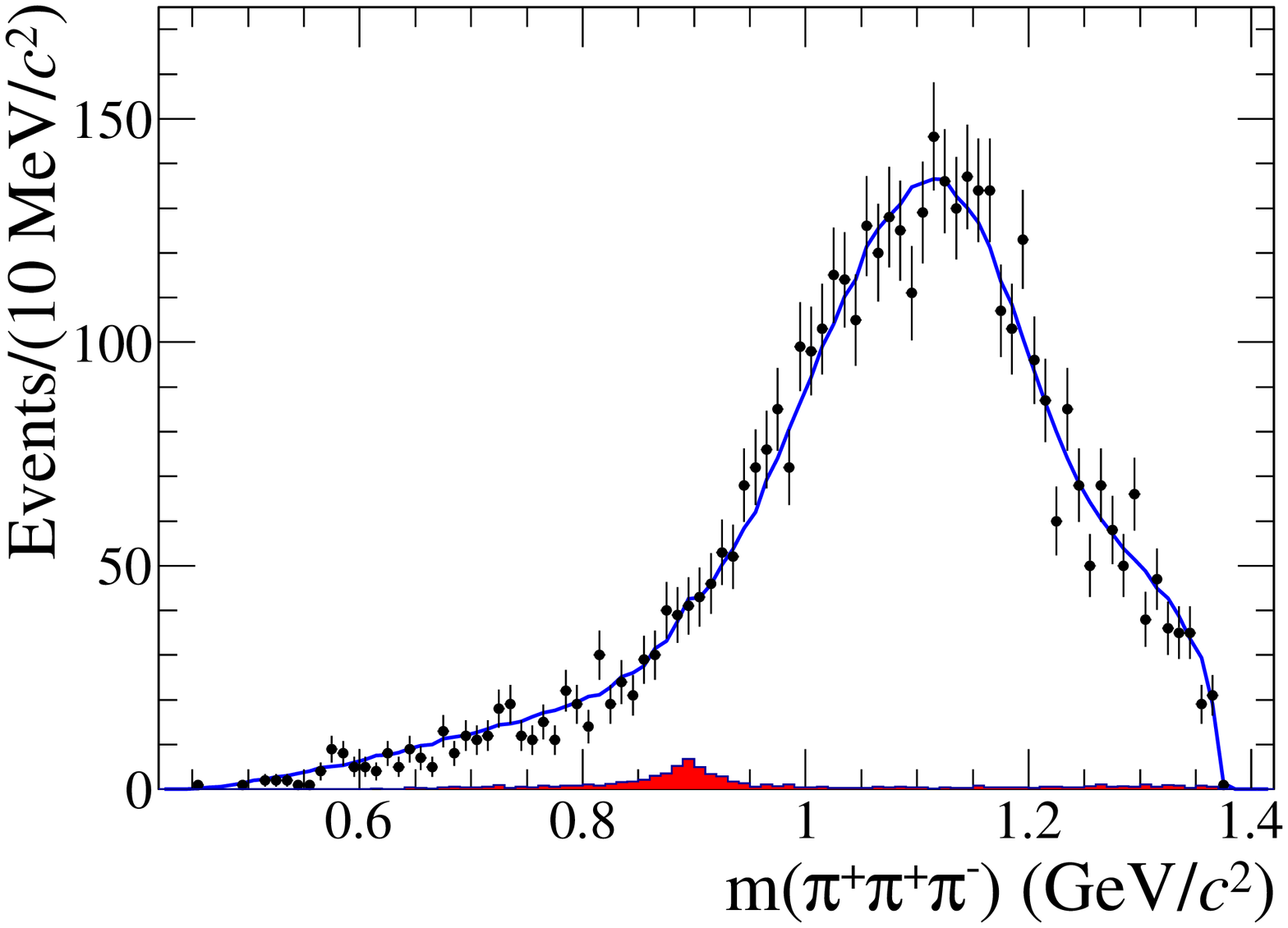}
\put(-25,90){(h)}
\end{minipage}
\begin{minipage}[b]{0.28\textwidth}
\epsfig{width=1.00\textwidth,clip=true,file=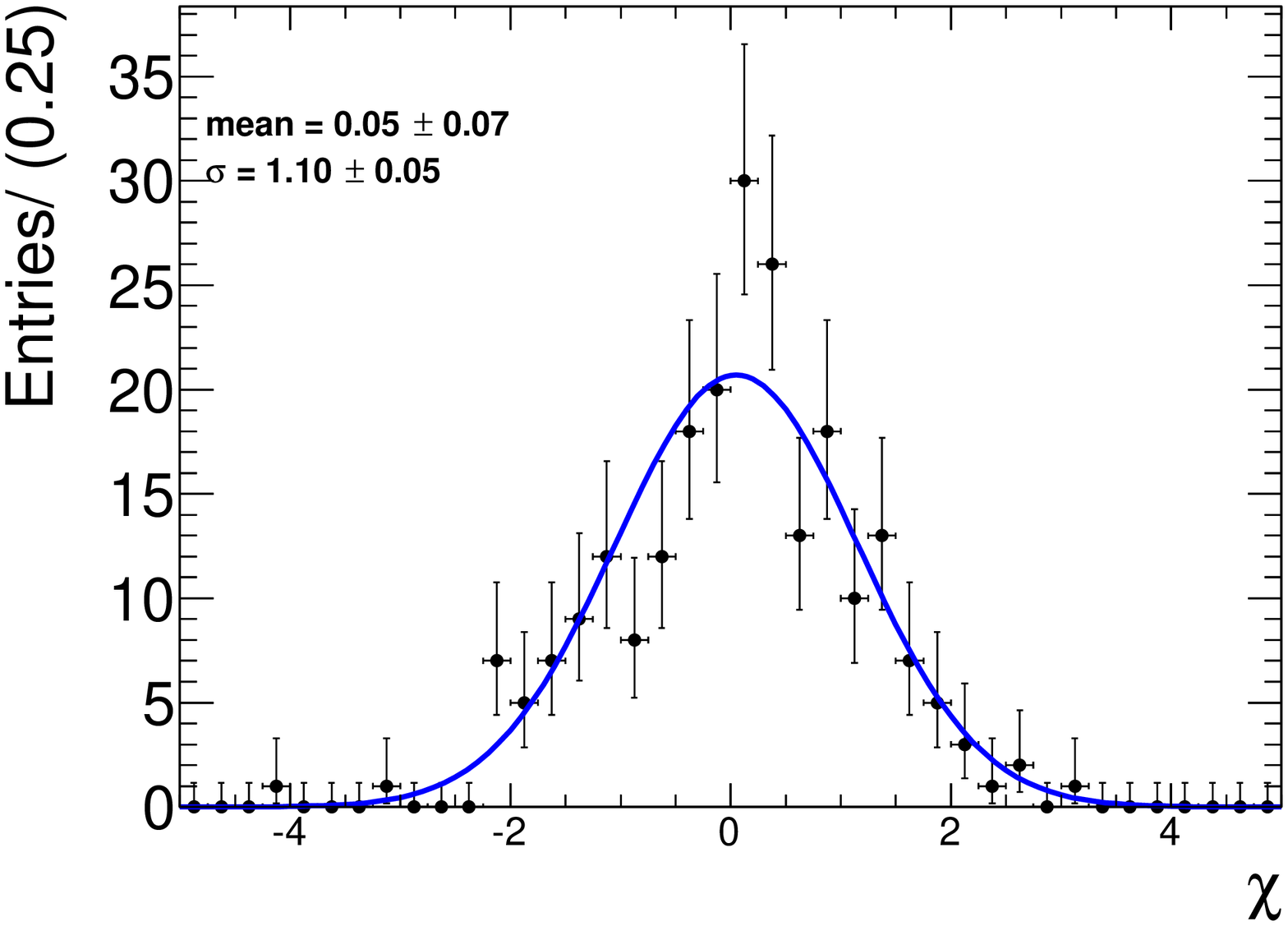}
\put(-25,90){(i)}
\end{minipage}
\caption{The projections of (a) $K_{S}^{0}\pi^{-}$, (b) $K_{S}^{0}\pi^{+}_{1}$, (c) $K_{S}^{0}\pi^{+}_{2}$,
(d) $\pi^{+}_{1}\pi^{-}$, (e) $\pi^{+}_{2}\pi^{-}$, (f) $K_{S}^{0}\pi^{+}_{1}\pi^{-}$, (g) $K_{S}^{0}\pi^{+}_{2}\pi^{-}$, and
(h) $\pi^{+}\pi^{+}\pi^{-}$ invariant mass spectra, where the dots with error are data, and the curves are the fit projections. 
The small red (colors online) histogram in each projection shows the $D^{+} \rightarrow K_{S}^{0}K_{S}^{0}\pi^{+}$
peaking background. The dip around the $K_{S}^{0}$ peak comes from the used  requirement to suppress 
the $D^{+} \rightarrow K_{S}^{0}K_{S}^{0}\pi^{+}$ peaking background. 
The identical pions are sorted with same method mentioned in Sec.~\ref{Goodness of Fit}. 
Figure (i) shows the fit (curve) to the distribution of $\chi$ (points with error bars)
with a Gaussian function and the fitted values of the parameters (mean and width of Gaussian).}
\label{fig:proj}
\end{figure*}

Finally, our nominal fit model includes 13 amplitudes, in which 8 of them can be summarized 
into four different components.
The nominal fit yields a goodness of fit value of $\chi^{2}/\nu = 275.15 / 204 = 1.35$. 
The projections of the invariant mass spectra and the distribution of $\chi$ are shown in Fig.~\ref{fig:proj}.
All the amplitudes and the corresponding significances and phases, as well as the FFs of amplitudes and
components are listed in Table ~\ref{tab:amps}, where the last row 
of each box is the coherent sum of earlier published amplitudes (components).
For the phases and FFs, the first and second uncertainties are statistical and systematic, respectively.
The systematic uncertainties are discussed below.
Other tested amplitudes when determining the nominal fit model, but finally not used, are listed in Appendix A.

\begin{table*}[hbtp]
\begin{center}
\caption{Significances and phases for different amplitudes, 
as well as FFs for amplitudes and components (the last row of each box), where 
the first and second uncertainties are statistical and systematic, respectively.
The $f_{0}(500)$ and $\rho^{0}$ resonances decay to $\pi^{+}\pi^{-}$,
and the $K^{*-}$ resonance decays to $K_{S}^{0}\pi^{-}$.}
\begin{tabular}{lccc} \hline
 Amplitude                                      & Significance ($\sigma$) & Phase & FF\\ \hline
$D^{+}\rightarrow K_{S}^{0}a_{1}(1260)^{+}(\rho^{0}\pi^{+}[S])$    &$>10$& 0.0 (fixed)           & $0.384\pm0.021\pm0.029$\\          
$D^{+}\rightarrow K_{S}^{0}a_{1}(1260)^{+}(\rho^{0}\pi^{+}[D])$    & $4.3$ & $-1.55\pm0.16\pm0.22$ & $0.004\pm0.002\pm0.001$\\          
$D^{+}\rightarrow K_{S}^{0}a_{1}(1260)^{+}(\rho^{0}\pi^{+})$       &   -   &          -            & $0.403\pm0.021\pm0.029$\\\hline    %
$D^{+}\rightarrow K_{S}^{0}a_{1}(1260)^{+}(f_{0}(500)\pi^{+})$     &$>10$& $-1.82\pm0.08\pm0.10$ & $0.055\pm0.007\pm0.017$\\\hline    
$D^{+}\rightarrow \bar{K}_{1}(1400)^{0}(K^{*-}\pi^{+}[S])\pi^{+}$  &$>10$& $-2.68\pm0.05\pm0.07$ & $0.221\pm0.012\pm0.018$\\          
$D^{+}\rightarrow \bar{K}_{1}(1400)^{0}(K^{*-}\pi^{+}[D])\pi^{+}$  &$>10$& $-2.24\pm0.10\pm0.07$ & $0.015\pm0.002\pm0.001$\\          
$D^{+}\rightarrow \bar{K}_{1}(1400)^{0}(K^{*-}\pi^{+})\pi^{+}$     &   -   &          -            & $0.216\pm0.012\pm0.011$\\\hline    %
$D^{+}\rightarrow \bar{K}_{1}(1270)^{0}(K_{S}^{0}\rho^{0}[S])\pi^{+}$&$9.7$& $-0.56\pm0.09\pm0.11$ & $0.024\pm0.003\pm0.007$\\\hline    
$D^{+}\rightarrow \bar{K}(1460)^{0}(K^{*-}\pi^{+})\pi^{+}$         &$>10$  & $-2.50\pm0.07\pm0.06$ & $0.068\pm0.006\pm0.002$\\\hline    
$D^{+}\rightarrow \bar{K}(1460)^{0}(K_{S}^{0}\rho^{0})\pi^{+}$     &$6.1$  & $-2.65\pm0.18\pm0.25$ & $0.008\pm0.002\pm0.005$\\\hline    
$D^{+}\rightarrow \bar{K}_{1}(1650)^{0}(K^{*-}\pi^{+}[S])\pi^{+}$  &$6.5$  & $ 0.95\pm0.14\pm0.22$ & $0.016\pm0.004\pm0.009$\\\hline    
$D^{+}\rightarrow (K_{S}^{0}\rho^{0}[S])_{A}\pi^{+}$               &$>10$  & $-1.88\pm0.08\pm0.05$ & $0.057\pm0.007\pm0.023$\\          
$D^{+}\rightarrow (K_{S}^{0}\rho^{0}[D])_{A}\pi^{+}$               &$7.0$  & $ 2.77\pm0.12\pm0.14$ & $0.008\pm0.002\pm0.001$\\          
$D^{+}\rightarrow (K_{S}^{0}\rho^{0})_{A}\pi^{+}$                  &   -   &          -            & $0.064\pm0.007\pm0.030$\\\hline    %
$D^{+}\rightarrow (K_{S}^{0}(\pi^{+}\pi^{-})_{S})_{A}\pi^{+}$      &$>10$& $-3.08\pm0.06\pm0.04$ & $0.064\pm0.005\pm0.007$\\          
$D^{+}\rightarrow ((K_{S}^{0}\pi^{+})_{S-{\rm wave}}\pi^{-})_{P}\pi^{+}$ &$>10$& $ 2.10\pm0.08\pm0.28$ & $0.017\pm0.003\pm0.004$\\          
$D^{+}\rightarrow K_{S}^{0}\pi^{+}\pi^{+}\pi^{-}$ non-resonance    &   -   &          -            & $0.081\pm0.006\pm0.006$\\\hline    
\end{tabular}
\label{tab:amps}
\end{center}
\end{table*}

\section{Systematic Uncertainties}
\label{p_sys_unc}
The systematic uncertainties are categorized into the following sources:
(I) masses and widths of the intermediate resonances, (II) effective radius of intermediate resonances and $D^{+}$,
(III) parameters in $K_{S}^{0}\pi^{+}$ $S$-wave parameterization, (IV) parameters in $\rho-\omega$ mixing parameterization, (V) line shape of $f_{0}(500)$,
(VI) effect from peaking background, (VII) effect from general background, and (VIII) fit procedure.  
The systematic uncertainties of the phases of amplitudes and the FFs of amplitudes and components due to different contributions
are given in Tables~\ref{Tab: sys_phi} and~\ref{Tab: sys_FF}, respectively.
These uncertainties are given in units of standard deviations $\sigma_{{\rm stat}}$ and
are added in quadrature to obtain the total systematic uncertainties, as they are uncorrelated. 
\begin{table*}[hbtp]
\begin{center}
\caption{Systematic uncertainties of phases for amplitudes. The different sources include 
(I) masses and widths of the intermediate resonances, (II) effective radius of intermediate resonances and $D^{+}$,
(III) parameters in the $K_{S}^{0}\pi^{+}$ $S$-wave parameterization, (IV) parameters in the $\rho-\omega$ mixing parameterization, (V) line shape of the $f_{0}(500)$,
(VI) effect from peaking background, (VII) effect from general background, and (VIII) fit procedure.}
\begin{tabular}{lccccccccc} \hline
\multirow{2}{*}{Amplitude}                                        &\multicolumn{8}{c}{Source ($\sigma_{{\rm stat}}$)}& $~$  \\
$~$                                                                & I     & II  & III & IV  &  V  & VI  & VII &VIII & total \\ \hline
$D^{+}\rightarrow K_{S}^{0}a_{1}(1260)^{+}(\rho^{0}\pi^{+}[D])$    &0.317  &0.413&1.221&0.059&0.273&0.042&0.057&0.061&1.412\\
$D^{+}\rightarrow K_{S}^{0}a_{1}(1260)^{+}(f_{0}(500)\pi^{+})$     &0.265  &0.343&1.110&0.262&  -  &0.220&0.058&0.071&1.243\\
$D^{+}\rightarrow \bar{K}_{1}(1400)^{0}\pi^{+}(K^{*-}\pi^{+}[S])$  &0.872  &0.362&1.006&0.131&0.257&0.003&0.051&0.058&1.412\\
$D^{+}\rightarrow \bar{K}_{1}(1400)^{0}\pi^{+}(K^{*-}\pi^{+}[D])$  &0.393  &0.252&0.451&0.068&0.062&0.001&0.097&0.149&0.679\\
$D^{+}\rightarrow \bar{K}_{1}(1270)^{0}\pi^{+}(K_{S}^{0}\rho^{0}[S])$&1.135&0.349&0.123&0.021&0.012&0.131&0.121&0.121&1.213\\
$D^{+}\rightarrow \bar{K}(1460)^{0}(K^{*-}\pi^{+})\pi^{+}$         &0.786  &0.032&0.152&0.049&0.128&0.028&0.092&0.054&0.820\\
$D^{+}\rightarrow \bar{K}(1460)^{0}(K_{S}^{0}\rho^{0})\pi^{+}$     &0.573  &0.022&1.249&0.023&0.261&0.070&0.062&0.139&1.409\\
$D^{+}\rightarrow \bar{K}_{1}(1650)^{0}\pi^{+}(K^{*-}\pi^{+}[S])$  &1.171  &0.166&0.948&0.026&0.089&0.066&0.118&0.051&1.526\\
$D^{+}\rightarrow (K_{S}^{0}\rho^{0}[S])_{A}\pi^{+}$               &0.539  &0.307&0.217&0.015&0.061&0.007&0.115&0.050&0.672\\
$D^{+}\rightarrow (K_{S}^{0}\rho^{0}[D])_{A}\pi^{+}$               &0.173  &0.278&1.057&0.038&0.273&0.045&0.057&0.100&1.147\\
$D^{+}\rightarrow (K_{S}^{0}(\pi^{+}\pi^{-})_{S})_{A}\pi^{+}$      &0.254  &0.508&0.442&0.072&0.010&0.058&0.092&0.050&0.733\\
$D^{+}\rightarrow ((K_{S}^{0}\pi^{+})_{S-{\rm wave}}\pi^{-})_{P}\pi^{+}$ &0.142  &0.226&3.309&0.083&0.192&0.027&0.059&0.125&3.330\\
\hline
\end{tabular}
\label{Tab: sys_phi}
\caption{Systematic uncertainties of FFs for amplitudes and components.
The different sources include
(I) masses and widths of the intermediate resonances, (II) effective radius of intermediate resonances and $D^{+}$,
(III) parameters in the $K_{S}^{0}\pi^{+}$ $S$-wave parameterization, (IV) parameters in $\rho-\omega$ mixing parameterization, (V) line shape of the $f_{0}(500)$,
(VI) effect from peaking background, (VII) effect from general background, and (VIII) fit procedure.}
\begin{tabular}{lccccccccc} \hline
\multirow{2}{*}{Amplitude and component}                         &\multicolumn{8}{c}{Source ($\sigma_{{\rm stat}}$)}& $~$ \\
$~$                                                                & I     & II  & III & IV  &  V  & VI  & VII &VIII & total \\ \hline
$D^{+}\rightarrow K_{S}^{0}a_{1}(1260)^{+}(\rho^{0}\pi^{+}[S])$    &0.299  &0.831&0.496&0.069&0.877&0.215&0.023&0.143&1.367\\      
$D^{+}\rightarrow K_{S}^{0}a_{1}(1260)^{+}(\rho^{0}\pi^{+}[D])$    &0.137  &0.335&0.032&0.078&0.014&0.028&0.054&0.085&0.386\\      
$D^{+}\rightarrow K_{S}^{0}a_{1}(1260)^{+}(\rho^{0}\pi^{+})$       &0.301  &0.885&0.529&0.054&0.870&0.217&0.014&0.125&1.406\\      %
$D^{+}\rightarrow K_{S}^{0}a_{1}(1260)^{+}(f_{0}(500)\pi^{+})$     &0.534  &0.538&2.369&0.050&0.553&0.215&0.097&0.085&2.410\\      
$D^{+}\rightarrow \bar{K}_{1}(1400)^{0}\pi^{+}(K^{*-}\pi^{+}[S])$  &1.260  &0.094&0.306&0.003&0.093&0.177&0.174&0.060&1.328\\      
$D^{+}\rightarrow \bar{K}_{1}(1400)^{0}\pi^{+}(K^{*-}\pi^{+}[D])$  &0.286  &0.099&0.216&0.007&0.041&0.027&0.042&0.078&0.386\\      
$D^{+}\rightarrow \bar{K}_{1}(1400)^{0}\pi^{+}(K^{*-}\pi^{+})$     &0.857  &0.078&0.221&0.002&0.066&0.123&0.119&0.063&0.910\\      %
$D^{+}\rightarrow \bar{K}_{1}(1270)^{0}\pi^{+}(K_{S}^{0}\rho^{0}[S])$&1.151&0.274&1.511&0.071&0.480&0.172&0.061&0.086&1.990\\      
$D^{+}\rightarrow \bar{K}(1460)^{0}(K^{*-}\pi^{+})\pi^{+}$         &0.288  &0.081&0.162&0.001&0.048&0.016&0.016&0.071&0.351\\      
$D^{+}\rightarrow \bar{K}(1460)^{0}(K_{S}^{0}\rho^{0})\pi^{+}$     &0.365  &0.546&2.288&0.044&0.374&0.194&0.153&0.058&2.423\\      
$D^{+}\rightarrow \bar{K}_{1}(1650)^{0}\pi^{+}(K^{*-}\pi^{+}[S])$  &1.836  &0.862&0.077&0.007&0.164&0.095&0.195&0.063&2.049\\      
$D^{+}\rightarrow (K_{S}^{0}\rho^{0}[S])_{A}\pi^{+}$               &0.644  &0.758&3.139&0.036&0.124&0.154&0.037&0.058&3.300\\      
$D^{+}\rightarrow (K_{S}^{0}\rho^{0}[D])_{A}\pi^{+}$               &0.188  &0.248&0.334&0.044&0.010&0.072&0.001&0.092&0.474\\      
$D^{+}\rightarrow (K_{S}^{0}\rho^{0})_{A}\pi^{+}$                  &0.863  &0.876&4.287&0.031&0.131&0.236&0.066&0.078&4.469\\      %
$D^{+}\rightarrow (K_{S}^{0}(\pi^{+}\pi^{-})_{S})_{A}\pi^{+}$      &0.751  &0.318&0.933&0.035&0.243&0.548&0.363&0.149&1.432\\      
$D^{+}\rightarrow ((K_{S}^{0}\pi^{+})_{S-{\rm wave}}\pi^{-})_{P}\pi^{+}$ &0.347  &0.073&1.422&0.014&0.128&0.259&0.039&0.086&1.497\\      
$D^{+}\rightarrow K_{S}^{0}\pi^{+}\pi^{+}\pi^{-}$ non-resonance    &0.604  &0.256&0.191&0.025&0.153&0.580&0.327&0.078&0.969\\      %
\hline
\end{tabular}
\label{Tab: sys_FF}
\end{center}
\end{table*}

To estimate the systematic uncertainties, the fit is altered to investigate the effect from each source.
For the masses and widths of the intermediate resonances given by the PDG~\cite{PDG},
they are shifted within the uncertainties from the PDG~\cite{PDG}.
The masses and widths of $a_{1}(1260)^{+}$ and $\bar{K}(1460)$, 
as well as the relative magnitude and phase of $\omega$ in $\rho-\omega$ parameterization
are shifted within the uncertainties given by the likelihood scans.
The barrier effective radius R is varied within $\pm1 {\mbox{\,GeV}}^{-1}$.
The input parameters of $K_{S}^{0}\pi^{+}$ $S$-wave model are varied within their uncertainties given by Ref.~\cite{KPiS}.
For the resonance $f_{0}(500)$, the propagator is replaced by RBW function with mass and width fixed at
526 MeV/$c^2$ and 535 MeV~\cite{Ablikim:2004qna}, respectively.
Since different propagators have different normalization factors,
for the amplitude with $f_{0}(500)$ involved, the shift effects on the FF are only considered.
The effect from the peaking background $D^{+} \rightarrow K^{0}_{S}K^{0}_{S}\pi^{+}$ is estimated 
by altering the number of background events to be half of it in the nominal fit.
The uncertainty from general background is studied by taking the background events into account, which are estimated from 
the average $M_{{\rm BC}}$ ($(M_{{\rm BC}}(D_{{\rm tag}}) + M_{{\rm BC}}(D_{{\rm signal}}))/2$) sideband region of $[1.830,\,1.858]$ GeV$/c^{2}$. 
Individual changes of the results with respect to the 
nominal one are taken as the corresponding systematic uncertainties.

To evaluate the uncertainty from the fit procedure, 
we generate 300 sets of signal MC samples according to the nominal results in this analysis. 
Each sample, which has equivalent size as the data, is analyzed with the same method as data analysis.
We fit the resulting pull distributions, $\frac{V_{{\rm input}}-V_{{\rm fit}}}{\sigma_{{\rm fit}}}$, 
where $V_{{\rm input}}$ is the input value
in the generator, and $V_{{\rm fit}}$ and $\sigma_{{\rm fit}}$ are the output value 
and the corresponding statistical uncertainty, respectively. 
Fits to the pull distributions with Gaussian functions show
no obvious biases and under- or over-estimations on statistical uncertainties.
We add in quadrature the mean and the mean error
of the pull and multiply this number with the statistical error to get the systematic error.
The results are given in Table~\ref{Pull}, in which the corresponding uncertainties 
are the statistical uncertainties of the respective fits.
\begin{table*}[hbtp]
\begin{center}
\caption{Mean and width of the pull distributions for phases and FFs with statistical uncertainties.}
\begin{tabular}{lcccc} \hline
\multirow{2}{*}{Amplitude and component}                         &\multicolumn{2}{c}{Phase} & \multicolumn{2}{c}{FF} \\
$~$                                                                &  mean         & width       &      mean    & width  \\ \hline
$D^{+}\rightarrow K_{S}^{0}a_{1}(1260)^{+}(\rho^{0}\pi^{+}[S])$    &  -            &    -        &$-0.13\pm0.06$&$0.96\pm0.04$\\ 
$D^{+}\rightarrow K_{S}^{0}a_{1}(1260)^{+}(\rho^{0}\pi^{+}[D])$    &$0.01\pm0.06$  &$1.01\pm0.04$&$0.06\pm0.06$ &$0.96\pm0.04$\\ 
$D^{+}\rightarrow K_{S}^{0}a_{1}(1260)^{+}(\rho^{0}\pi^{+})$       &  -            &    -        &$-0.11\pm0.06$&$0.97\pm0.04$\\ %
$D^{+}\rightarrow K_{S}^{0}a_{1}(1260)^{+}(f_{0}(500)\pi^{+})$     &$0.05\pm0.05$  &$0.89\pm0.04$&$0.06\pm0.06$ &$1.01\pm0.04$\\ 
$D^{+}\rightarrow \bar{K}_{1}(1400)^{0}\pi^{+}(K^{*-}\pi^{+}[S])$  &$-0.03\pm0.05$ &$0.92\pm0.04$&$0.00\pm0.06$ &$1.03\pm0.04$\\ 
$D^{+}\rightarrow \bar{K}_{1}(1400)^{0}\pi^{+}(K^{*-}\pi^{+}[D])$  &$0.14\pm0.05$  &$0.93\pm0.04$&$0.05\pm0.06$ &$0.97\pm0.04$\\ 
$D^{+}\rightarrow \bar{K}_{1}(1400)^{0}\pi^{+}(K^{*-}\pi^{+})$     &  -            &    -        &$0.02\pm0.06$ &$0.97\pm0.04$\\ %
$D^{+}\rightarrow \bar{K}_{1}(1270)^{0}\pi^{+}(K_{S}^{0}\rho^{0}[S])$&$0.11\pm0.05$&$0.95\pm0.04$&$-0.07\pm0.05$&$0.95\pm0.04$\\ 
$D^{+}\rightarrow \bar{K}(1460)^{0}(K^{*-}\pi^{+})\pi^{+}$         &$-0.02\pm0.05$ &$0.91\pm0.04$&$0.05\pm0.05$ &$0.95\pm0.04$\\ 
$D^{+}\rightarrow \bar{K}(1460)^{0}(K_{S}^{0}\rho^{0})\pi^{+}$     &$0.13\pm0.05$  &$0.94\pm0.04$&$0.03\pm0.05$ &$0.95\pm0.04$\\ 
$D^{+}\rightarrow \bar{K}_{1}(1650)^{0}\pi^{+}(K^{*-}\pi^{+}[S])$  &$0.01\pm0.05$  &$0.93\pm0.04$&$-0.02\pm0.06$&$1.01\pm0.04$\\ 
$D^{+}\rightarrow (K_{S}^{0}\rho^{0}[S])_{A}\pi^{+}$               &$0.00\pm0.05$  &$0.93\pm0.04$&$-0.03\pm0.05$&$0.89\pm0.04$\\ 
$D^{+}\rightarrow (K_{S}^{0}\rho^{0}[D])_{A}\pi^{+}$               &$-0.08\pm0.06$ &$1.06\pm0.04$&$0.07\pm0.06$ &$1.06\pm0.04$\\ 
$D^{+}\rightarrow (K_{S}^{0}\rho^{0})_{A}\pi^{+}$                  &  -            &    -        &$0.06\pm0.05$ &$0.93\pm0.04$\\ %
$D^{+}\rightarrow (K_{S}^{0}(\pi^{+}\pi^{-})_{S})_{A}\pi^{+}$      &$0.00\pm0.05$  &$0.87\pm0.04$&$-0.14\pm0.05$&$0.92\pm0.04$\\ 
$D^{+}\rightarrow ((K_{S}^{0}\pi^{+})_{S-{\rm wave}}\pi^{-})_{P}\pi^{+}$ &$0.11\pm0.06$  &$0.97\pm0.04$&$0.07\pm0.05$ &$0.93\pm0.04$\\ 
$D^{+}\rightarrow K_{S}^{0}\pi^{+}\pi^{+}\pi^{-}$ non-resonance    &  -            &    -        &$-0.06\pm0.05$&$0.95\pm0.04$\\ %
\hline
\end{tabular}
\label{Pull}
\end{center}
\end{table*}

The effects from the tracking/PID efficiency difference between data and MC simulation, 
as well as the resolution are also investigated. 
For tracking/PID efficiency, a factor related to the correction is considered 
when calculating the normalization integral of Eq.~(\ref{signalPDF}). 
The difference between the alternative fit and the nominal fit is found to be negligible. 
The effect from the resolution is estimated from the difference of the pull distribution 
obtained from these 300 sets of signal MC samples using the generated and reconstructed four-momenta, 
which is also found to be negligible.

\section{Conclusion}
\label{CONLUSION}
We have determined the intermediate state contributions to the decay $D^{+}\rightarrow K_{S}^{0}\pi^{+}\pi^{+}\pi^{-}$ 
from an amplitude analysis. With the fit fraction of the $n^{{\rm th}}$ component ${\rm FF}(n)$ obtained from this analysis, 
we calculate the corresponding BF:
$\mathcal{B}(n) = \mathcal{B}(D^{+}\rightarrow K_{S}^{0}\pi^{+}\pi^{+}\pi^{-}) \times {\rm FF}(n)$,
where $\mathcal{B}(D^{+}\rightarrow K_{S}^{0}\pi^{+}\pi^{+}\pi^{-}) = (2.97\pm0.11)\%$ is the
total inclusive BF quoted from
the PDG~\cite{PDG}.
The results on the BFs are shown in Table~\ref{tab:final res_BR}.
\begin{table}[htbp]
\caption{The results of BFs for different components.
The first, second and third errors are statistical, systematical and the uncertainty related to
$\mathcal{B}(D^{+} \rightarrow K_{S}^{0}\pi^{+}\pi^{+}\pi^{-})$~\cite{PDG}, respectively. The 
$f_{0}(500)$ and $\rho^{0}$ resonances decay to $\pi^{+}\pi^{-}$,
and the $K^{*-}$ resonance decays to $K_{S}^{0}\pi^{-}$.}
\begin{center}
\footnotesize
\begin{tabular}{l|c} \hline
Component & Branching fraction~(\%) \\ \hline
$D^{+}\rightarrow K_{S}^{0}a_{1}(1260)^{+}(\rho^{0}\pi^{+})$      & $1.197\pm0.062\pm0.086\pm0.044$\\
$D^{+}\rightarrow K_{S}^{0}a_{1}(1260)^{+}(f_{0}(500)\pi^{+})$    & $0.163\pm0.021\pm0.005\pm0.006$\\
$D^{+}\rightarrow \bar{K}_{1}(1400)^{0}(K^{*-}\pi^{+})\pi^{+}$    & $0.642\pm0.036\pm0.033\pm0.024$\\
$D^{+}\rightarrow \bar{K}_{1}(1270)^{0}(K_{S}^{0}\rho^{0})\pi^{+}$& $0.071\pm0.009\pm0.021\pm0.003$\\
$D^{+}\rightarrow \bar{K}(1460)^{0}(K^{*-}\pi^{+})\pi^{+}$        & $0.202\pm0.018\pm0.006\pm0.007$\\
$D^{+}\rightarrow \bar{K}(1460)^{0}(K_{S}^{0}\rho^{0})\pi^{+}$    & $0.024\pm0.006\pm0.015\pm0.009$\\
$D^{+}\rightarrow \bar{K}_{1}(1650)^{0}(K^{*-}\pi^{+})\pi^{+}$    & $0.048\pm0.012\pm0.027\pm0.002$\\
$D^{+}\rightarrow K_{S}^{0}\pi^{+}\rho^{0}$                       & $0.190\pm0.021\pm0.089\pm0.007$\\
$D^{+} \rightarrow K_{S}^{0}\pi^{+}\pi^{+}\pi^{-}$                & $0.241\pm0.018\pm0.018\pm0.009$\\
\hline
\end{tabular}
\label{tab:final res_BR}
\end{center}
\end{table}

Compared with the previous measurements~\cite{MarKIII}, the precisions of the sub decay modes are significantly improved. The dominant
intermediate process is $D^{+}\rightarrow K_{S}^{0}a_{1}(1260)^{+}(\rho^{0}\pi^{+})$, which agrees with the measurement of Mark~III~\cite{MarKIII}.
We also extract the BFs of $D^{+}\rightarrow K_{S}^{0}a_{1}(1260)^{+}(f_{0}(500)\pi^{+})$,
$D^{+}\rightarrow \bar{K}_{1}(1400)^{0}(K^{*-}\pi^{+})\pi^{+}$, and
$D^{+}\rightarrow \bar{K}_{1}(1270)^{0}(K_{S}^{0}\rho^{0})\pi^{+}$ decays for the first time. 
Comparing with the decay of  $D^{0} \rightarrow K^{-}\pi^{+}\pi^{+}\pi^{-}$~\cite{Ablikim:2017eqz,Aaij:2017kbo}, the decay mode
$D \to K a_{1}(1260)$ is found to be the dominant substructure in both $D^{0}$ and $D^{+}$ decays.
For the two $K_{1}$ states, the contributions from $D \to K_{1}(1270) \pi$ is at the same level for both $D^{+}$ and $D^{0}$ decays. 
For $D \to K_{1}(1400) \pi$, the related BF in $D^{+}$ decays is found to be greater than that in $D^{0}$ decay by one order of magnitude. 
These results provide criteria to further investigate the mixture between these two axial-vector kaon states~\cite{Cheng:2003bn,PRD81074031,HYCheng}.

\begin{acknowledgements}
\label{sec:acknowledgement}
\vspace{-0.4cm}
The BESIII collaboration thanks the staff of BEPCII and the IHEP computing center for their strong support. 
This work is supported in part by National Key Basic Research Program of China under Contract No. 2015CB856700; 
National Natural Science Foundation of China (NSFC) under Contracts Nos. 11075174, 11121092, 11425524, 11475185, 11625523, 11635010, 11735014; 
the Chinese Academy of Sciences (CAS) Large-Scale Scientific Facility Program; 
the CAS Center for Excellence in Particle Physics (CCEPP); 
Joint Large-Scale Scientific Facility Funds of the NSFC and CAS under Contracts Nos. U1532257, U1532258, U1732263; 
CAS Key Research Program of Frontier Sciences under Contracts Nos. QYZDJ-SSW-SLH003, QYZDJ-SSW-SLH040; 
100 Talents Program of CAS; INPAC and Shanghai Key Laboratory for Particle Physics and Cosmology; '
German Research Foundation DFG under Contract No. Collaborative Research Center CRC 1044; 
Istituto Nazionale di Fisica Nucleare, Italy; Koninklijke Nederlandse Akademie van Wetenschappen (KNAW) under Contract No. 530-4CDP03; 
Ministry of Development of Turkey under Contract No. DPT2006K-120470; 
National Science and Technology fund; The Swedish Research Council; The Knut and Alice Wallenberg Foundation;
U. S. Department of Energy under Contracts Nos. DE-FG02-05ER41374, DE-SC-0010118, DE-SC-0010504, DE-SC-0012069; University of Groningen (RuG) 
and the Helmholtzzentrum fuer Schwerionenforschung GmbH (GSI), Darmstadt.
\end{acknowledgements}

\section{Appendix A: Amplitudes Tested}
\label{sec:appenA}
We list the amplitudes which are tested when searching for 
the nominal fit model but not used in the final result due to the low significance ($<5\sigma$).\\
{\footnotesize \bf Amplitudes with excited states ($m>1.0$ GeV$/c^{2}$) involved}\\
{\footnotesize
$D^{+} \rightarrow \bar{K}_{1}(1270)^{0}\pi^{+}$, $\bar{K}_{1}(1270)^{0} \rightarrow K_{S}^{0} \rho^{0}[D]$.\\
$D^{+} \rightarrow \bar{K}_{1}(1270)^{0}\pi^{+}$, $\bar{K}_{1}(1270)^{0} \rightarrow K^{*-} \pi^{+}[S,D]$.\\
$D^{+} \rightarrow K_{S}^{0} a_{2}(1320)^{+}$,
$a_{2}(1320)^{+} \rightarrow \rho^{0}\pi^{+}$ or $(\pi^{+}\pi^{-})_{T}\pi^{+}$.\\
$D^{+} \rightarrow K_{S}^{0} \pi(1300)^{+}$,
$\pi(1300)^{+} \rightarrow \rho^{0}\pi^{+}$ or $(\pi^{+}\pi^{-})_{S}\pi^{+}$.\\
$D^{+} \rightarrow K_{S}^{0} a_{1}(1640)^{+}$,
$a_{1}(1640)^{+} \rightarrow \rho^{0}\pi^{+}[S,D]$ or $(\pi^{+}\pi^{-})_{S}\pi^{+}$.\\
$D^{+} \rightarrow \bar{K}(1460)^{0}\pi^{+}$,
$\bar{K}(1460)^{0} \rightarrow (K_{S}^{0}\pi^{-})_{S}\pi^{+}$.\\
$D^{+} \rightarrow \bar{K}_{2}(1580)^{0}\pi^{+}$,
$\bar{K}_{2}(1580)^{0} \rightarrow K^{*-}\pi^{+}$ or $(K_{S}^{0}\pi^{-})_{T}\pi^{+}$.\\
$D^{+} \rightarrow \bar{K}^{*}(1410)^{0}\pi^{+}$,
$\bar{K}^{*}(1410)^{0} \rightarrow K^{*-}\pi^{+}$ or $K_{S}^{0}\rho^{0}$.\\
}
{\footnotesize \bf Amplitudes with only $K^{*-}$, $\rho^{0}$ and $f_{0}(500)$ involved}\\
{\footnotesize
$D^{+} \rightarrow K^{*-}(\pi^{+}\pi^{-})_{S}$.\\
$D^{+} \rightarrow (K^{*-} \pi^{+})_{P,V,A,T}\pi^{+}$.\\
$D^{+} \rightarrow (K_{S}^{0}\rho^{0})_{V,T}\pi^{+}$.\\
$D^{+} \rightarrow K_{S}^{0}(\rho^{0}\pi^{+})_{P,V,A,T}$.\\
$D^{+} \rightarrow (K_{S}^{0}f_{0}(500))_{P,A,T}\pi^{+}$.\\
$D^{+} \rightarrow K_{S}^{0}(f_{0}(500)\pi^{+})_{P,A,T}$.\\
}
{\footnotesize \bf Amplitudes without resonant state involved}\\
{\footnotesize
$D^{+} \rightarrow (K_{S}^{0}(\pi^{+}\pi^{-})_{S})_{P,A,T}\pi^{+}$.\\
$D^{+} \rightarrow (K_{S}^{0}(\pi^{+}\pi^{-})_{V})_{P,V,A,T}\pi^{+}$.\\
$D^{+} \rightarrow (K_{S}^{0}(\pi^{+}\pi^{-})_{T})_{A,T}\pi^{+}$.\\
$D^{+} \rightarrow K_{S}^{0}((\pi^{+}\pi^{-})_{S}\pi^{+})_{P,A,T}$.\\
$D^{+} \rightarrow K_{S}^{0}((\pi^{+}\pi^{-})_{V}\pi^{+})_{P,V,A,T}$.\\
$D^{+} \rightarrow K_{S}^{0}((\pi^{+}\pi^{-})_{T}\pi^{+})_{A,T}$.\\
$D^{+} \rightarrow ((K_{S}^{0}\pi^{-})_{S-{\rm wave}}\pi^{+})_{A,T}\pi^{+}$.\\
$D^{+} \rightarrow ((K_{S}^{0}\pi^{-})_{V}\pi^{+})_{P,V,A,T}\pi^{+}$.\\
$D^{+} \rightarrow ((K_{S}^{0}\pi^{-})_{T}\pi^{+})_{A,T}\pi^{+}$.\\
}
{\footnotesize \bf Doubly Cabibbo-suppressed amplitudes}\\
{\footnotesize
$D^{+} \rightarrow K^{*+}\rho^{0}$.\\
$D^{+} \rightarrow K_{1}(1270)^{0}\pi^{+}$,
$K_{1}(1270)^{0} \rightarrow K^{*+}\pi^{-}[S,D]$.\\
}

\end{document}